\documentclass[
aps,%
12pt,%
final,%
notitlepage,%
oneside,%
onecolumn,%
nobibnotes,%
nofootinbib,
superscriptaddress,%
noshowpacs,%
centertags]%
{revtex4}
\begin{document}
%
%
\selectlanguage{english}
\title{Why Breakup of Photons and Pions
into Forward Dijets Is so Different: Predictions from Nonlinear
Nuclear $k_{\perp}$-factorization}
%
%
%
\author{N.N. Nikolaev}
\affiliation{IKP, Forschungszentrum J\"ulich, D-52425 J\"ulich,
Germany} \affiliation{L.D. Landau Institute for Theoretical
Physics, Chernogolovka, Russia}
\author{W. Sch\"afer}
\affiliation{IKP, Forschungszentrum J\"ulich, D-52425 J\"ulich,
Germany}
\author{B.G. Zakharov}
\affiliation{L.D. Landau Institute for Theoretical Physics,
Chernogolovka, Russia}
\author{V.R. Zoller}
\affiliation{Institute for Experimental and Theoretical Physics,
Moscow, Russia}

\maketitle
\begin{center}
\begin{minipage}{\textwidth - 2cm}
\small Based on an approach to non-Abelian propagation of color
dipoles in a nuclear medium we formulate a nonlinear
$k_{\perp}$-factorization for the breakup of photons and pions
into forward hard dijets in terms of the collective
Weizs\"acker-Williams (WW) glue of nuclei. We find quite distinct
practical consequences of nonlinear nuclear
$k_{\perp}$-factorization for interactions of pointlike photons
and non-pointlike pions. In the former case the large transverse
momentum $p_{\perp}$ of jets comes from the intrinsic  momentum of
quarks and antiquarks in the photon and nuclear effects manifest
themselves as an azimuthal decorrelation with an acoplanarity
momentum of the order of the nuclear saturation momentum $Q_A$. In
the breakup of pions off free nucleons to the leading order in
pQCD the spectator parton has a small transverse momentum and the
hard dijet cross section is suppressed. In the  breakup of pions
off heavy nuclei the forward hard jets are predicted to be
entirely decorrelated. We comment on the sensitivity of the pionic
dijet cross section to the pion distribution amplitude. The
predicted distinction between the breakup of photons and pions can
be tested by the sphericity and thrust analysis of the forward
hadronic system in the COMPASS experiment at CERN.
\end{minipage}
\end{center}
%
\newpage
\section*{Introduction}
The trademark of the conventional perturbative QCD (pQCD)
factorization theorems for hard interactions of leptons and
hadrons is that the hard scattering observables are linear
functionals of the appropriate parton densities in the projectile
and target \cite{Textbook}. In contrast to that, from the parton
model point of view the opacity of heavy nuclei to high energy
projectiles entails a highly nonlinear relationship between the
parton densities of free nucleons and nuclei. In deep inelastic
scattering (DIS) off nuclei there emerges a new large scale - the
nuclear saturation scale $Q_A$, - which separates the opaque
nucleus, i.e., nonlinear, and weak attenuation, i.e., linear,
regimes \cite{Mueller1,Mueller,McLerran,Saturation}. A priori it
is not obvious that in the nonlinear regime with the large
saturation scale one can define nuclear parton densities such that
they enter different observables in a universal manner, i.e., if
useful factorization theorems can be formulated for hard phenomena
in ultrarelativistic heavy ion collisions. In our previous work
\cite{Nonlinear,LIYaF} we presented a partial solution to this
problem - a nonlinear nuclear $k_{\perp}$-factorization for the
production of forward hard dijets in DIS off nuclei.

The salient feature of hard dijets in deep inelastic scattering
(DIS) and real photoabsorption is that the large transverse
momentum $p_{\perp}$ of forward jets comes from the intrinsic
momentum of quarks and antiquarks in the (virtual) photon. In the
$k_{\perp}$-factorization description of the underlying
photon-gluon fusion parton subprocess $\gamma^* g \to q\bar{q}$,
valid at small $x$, the disparity of the quark and antiquark
transverse momenta and departure from the exact back-to-back
configuration - the acoplanarity momentum - is caused by the
transverse momentum ${\boldmath{\kappa}}$ of gluons. It can be
quantified in terms of the unintegrated gluon density of the
target (see \cite{Azimuth,Forshaw} and references therein). Our
nonlinear $k_{\perp}$-factorization for breakup of photons into
dijets on nuclei gives a coherent description of the nuclear mass
number dependence of the dijet inclusive cross section in terms of
the collective Weizs\"acker-Williams (WW) unintegrated nuclear
glue. This WW nuclear glue has the form of an expansion over the
collective gluon structure function of spatially overlapping
nucleons \cite{NZfusion} of the Lorentz-contracted
ultrarelativistic nucleus \cite{NSSdijet,Saturation}. Apart from
the case of minijets with the $p_{\perp}$ comparable to or below
the saturation scale $Q_A$ the principal nuclear effect is a
broadening of the acoplanarity momentum distribution, and in
\cite{Nonlinear} we showed how this broadening can be calculated
through the collective WW nuclear unintegrated gluon density.

The breakup of pions into forward dijets in inelastic $\pi A$
collisions is an excitation of the quark-antiquark Fock states of
the pion. The intrinsic momentum of quarks in the non-pointlike
pion is limited. To the leading order in pQCD the breakup of the
pion goes via the pQCD analogue of the electrodisintegration of
the deuteron (for the review see \cite{Frullani}), i.e., the
subprocess $\pi g \to q\bar{q}$ in which the struck parton carries
the transverse momentum of the absorbed gluon and the spectator
parton emerges in the final state with the small transverse
momentum it had in the pion, fig. ~1. On the other hand, multiple
gluon exchange in collisions with opaque nuclei can give a large
transverse kick to both partons. In this communication we report
the nonlinear nuclear $k_{\perp}$-factorization formulas for the
breakup of pions into forward hard dijets off nuclei. One of our
central results is the prediction of a complete azimuthal
decorrelation of forward hard dijets.

At the parton level the produced forward jets retain the fraction
$z_{\pm}$ of the lightcone momentum of the pion they carried in
the incident pion. One may wonder whether the jet longitudinal
momentum distributions would  give a handle on the so-called pion
distribution amplitude ($\pi$DA) \cite{PionDA} which with some
reservations about large higher twist and next-to-leading order
pQCD corrections was indeed the case in the coherent diffractive
breakup of pions into djets off nuclei
\cite{NSSdijet,Chernyak,Regensburg}. Our analysis of nonlinear
nuclear $k_{\perp}$-factorization formulas shows that these
expectations are indeed met by the contribution to the dijet
inclusive cross section from the in-volume absorption of pions,
which comes from the perturbatively small $q\bar{q}$ dipole states
of the pion. However, this contribution is overwhelmed by a large
contribution from soft absorption of large $q\bar{q}$ dipole
states of the pion on the front face of a nucleus. In this case
there emerges some infrared-sensitive modulation of the
$z$-dependence of the $\pi$DA which brings a model-dependence into
tests of the pion wave function (WF).

The further presentation is organized as follows. The major thrust
is on the distinction between breakup of pointlike photons and
non-pointlike pions. To make the discussion self-contained we
present the basic formalism to a sufficient detail. In section 2
we set up the formalism with a brief discussion of the
decorrelation of jets in DIS and $\pi N$ scattering off free
nucleons. In section 3 we present the color-dipole $S$-matrix
formalism for the breakup into dijets on nuclear targets. In
section 4 we formulate a nonlinear nuclear
$k_{\perp}$-factorization for the inclusive dijet cross section in
terms of the collective WW unintegrated gluon density of the
nucleus and comment on the salient features of dijet production in
DIS off nuclei. The subject of section 5 is the breakup of
non-pointlike pions into dijets. In contrast to the breakup of
pointlike photons in DIS, excitation of two hard jets from pions
is only possible on heavy nuclei. The most striking difference
from DIS and real photoproduction is that the two pionic forward
hard jets produced off a nucleus are completely azimuthally
decorrelated. This leading contribution to the breakup cross
section comes from soft absorption of pions on the front face of a
nucleus. The hard contribution from the in-volume breakup gives
rise to back-to-back dijets as in DIS and has the form of a higher
twist correction. Its isolation is a challenging but not
impossible task and would allow the determination of the $\pi$DA.
In the Conclusions we summarize our principal findings and comment
on possible experimental tests of our predictions. The predicted
distinction between breakup of photons and pions can tested by the
sphericity and thrust analysis of the forward hadronic system in
the COMPASS experiment at CERN \cite{COMPASS}.

\section{Forward dijets off free nucleons and unintegrated glue
of the nucleon}

We set up the formalism on an example of breakup into dijets off
free nucleons. Production of high-mass forward hard dijets selects
excitation of the $q\bar{q}$ Fock states of the projectile photon
and meson. The relevant pQCD diagrams are shown in figs. 1a-1d. In
the color dipole approach
\cite{NZ91,NZ92,NZ94,NZZlett,NZglue,NZZdiffr,BGNPZunit,BGNPZshad}
the fundamental quantity is the total cross section for
interaction of the color dipole ${\bm{r}}$ with the target nucleon
\begin{eqnarray}
\sigma(x, {\bm{r}})&=& \alpha_S(r) \sigma_0(x)\int
d^2{\bm{\kappa}}
f({\mbox{\boldmath{$\kappa$}}} )\left[1 -\exp(i{\bm{\kappa}} {\bm{r}} )\right]\nonumber\\
& = & {1\over 2}\alpha_S(r) \sigma_0(x)\int
d^2{\mbox{\boldmath{$\kappa$}}} f({\mbox{\boldmath{$\kappa$}}}
)\left[1 -\exp(i{\mbox{\boldmath{$\kappa$}}} {\bm{r}} )\right]
\cdot \left[1 -\exp(-i{\mbox{\boldmath{$\kappa$}}} {\bm{r}}
)\right] \, , \label{eq:2.1}
\end{eqnarray}
where $f({\mbox{\boldmath{$\kappa$}}} )$  is normalized as $ \int
d^2{\mbox{\boldmath{$\kappa$}}}  f({\mbox{\boldmath{$\kappa$}}}
)=1 $ and is related to the so-called Fadin-Kuraev-Lipatov
unintegrated gluon density (\cite{BFKL}, for the recent review and
phenomenology see \cite{INDiffGlue,Andersson}) of the target
nucleon ${\cal F}(x,\kappa^2) = {\partial G(x,\kappa^2)/
\partial\log\kappa^2}$ by
\begin{eqnarray}
f({\mbox{\boldmath{$\kappa$}}} ) = {4\pi \over
N_c\sigma_0(x)}\cdot {1\over \kappa^4} \cdot {\cal
F}(x,\kappa^2)\, . \label{eq:2.2}
\end{eqnarray}
Here $\sigma_0(x)$ is an auxiliary soft parameter which drops out
from major nuclear observables.

First we consider DIS where the perturbative small size of dipoles
is set by the large virtuality $Q^2$ of the photon, and then
comment how the results extend to breakup of real photons and
pions into hard dijets. The total photoabsorption cross section
equals \begin{eqnarray} \sigma_N(Q^2,x) = \int d^2{\bm{r}} dz
|\Psi(Q^2,z,{\bm{r}})|^2 \sigma({\bm{r}})\, , \label{eq:2.3}
\end{eqnarray} where $\Psi(Q^2,z,{\bm{r}})= \langle  z,\bm{r}|\gamma^* \rangle$ is the WF of the $q\bar{q}$
Fock state of the photon, here-below we suppress the argument
$Q^{2}$ in $\Psi(Q^2,z,{\bm{r}})$. Upon the relevant Fourier
transformations one finds the momentum spectrum of the final state
(FS) quark prior the hadronization
\begin{eqnarray}
{d\sigma_N \over d^2{\bm{p}}_+ dz} = {\sigma_0(x)\over 2}\cdot {
\alpha_S({\bm{p}}_+^2) \over (2\pi)^2}
 \int d^2{\mbox{\boldmath{$\kappa$}}} f({\mbox{\boldmath{$\kappa$}}} )
\left|\langle \gamma^*|z,{\bm{p}}_+\rangle - \langle
\gamma^*|z,{\bm{p}}_+-{\mbox{\boldmath{$\kappa$}}}
\rangle\right|^2 \label{eq:2.4}
\end{eqnarray}
where ${\bm{p}}_+$ is the transverse momentum of the quark jet,
${\bm{p}}_- = -{\bm{p}}_+ + {\mbox{\boldmath{$\kappa$}}}$ is the
transverse momentum of the antiquark jet, $z_+= z$ and $ z_- =
1-z$ are the fractions of the photon's lightcone momentum carried
by the quark and antiquark jets, respectively. For our formalism
to apply we require that the variables $z_{\pm}$ for the observed
jets add up to unity, $x_{\gamma}=z_+ + z_- = 1$ and the rapidity
separation of jets is small, $z_+ \sim z_- \sim 1/2$, which in the
realm of DIS is often referred to as the unresolved or direct
photon interaction (\cite{ggFusion} and references therein).

Now notice that the transverse momentum of the gluon is precisely
the decorrelation momentum ${\bm{\Delta}} ={\bm{p}}_+ +{\bm{p}}_-$
so that in the still further differential form
\begin{eqnarray}
{d\sigma_N \over dz d^2{\bm{p}}_+ d^2{\bm{\Delta}}}& =&
{\sigma_0(x)\over 2}\cdot { \alpha_S({\bm{p}}^2) \over (2\pi)^2}
 f({\bm{\Delta}} )
\left|\langle \gamma^*|z,{\bm{p}}_+\rangle -
\langle \gamma^*|z,{\bm{p}}_+ -{\bm{\Delta}} \rangle\right|^2 \nonumber\\
& =& { \alpha_S({\bm{p}}^2) \over 2\pi N_c} \cdot
 { {\cal F}(x,{\bm{\Delta}}^2) \over \Delta^4}
\cdot \left|\langle \gamma^*|z,{\bm{p}}_+\rangle - \langle
\gamma^*|z,{\bm{p}}_+ -{\bm{\Delta}} \rangle\right|^2\, .
\label{eq:2.5}
\end{eqnarray}
This is the leading order result from $k_{\perp}$-factorization,
for the applications to DIS off free nucleons see
\cite{Azimuth,Forshaw} and references therein. Upon summing over
the helicities $\lambda,\overline{\lambda}$ of the final state
quark and antiquark, for transverse photons and flavor $f$ one has
\begin{eqnarray}
&&\left|\langle
\gamma^*|z,{\bm{p}}\rangle - \langle
\gamma^*|z,{\bm{p}}-{\mbox{\boldmath{$\kappa$}}} \rangle\right|^2_
{\lambda_{\gamma}=\pm 1} =
\nonumber\\
&&2N_c e_f^2\alpha_{em}\left\{ [z^{2}+(1-z)^{2}] \left({{\bm{p}}
\over {\bm{p}}^{2}+\varepsilon^{2}} -
{{\bm{p}}-{\mbox{\boldmath{$\kappa$}}} \over
({\bm{p}}-{\mbox{\boldmath{$\kappa$}}} )^{2}+\varepsilon^{2}}
\right)^2_{\lambda+\overline{\lambda}=0}\right.\nonumber\\
&&\left. +m_{f}^{2} \left({1   \over
{\bm{p}}^{2}+\varepsilon^{2}}- {1 \over
({\bm{p}}-{\mbox{\boldmath{$\kappa$}}}
)^{2}+\varepsilon^{2}}\right)^2
_{\lambda+\overline{\lambda}=\lambda_{\gamma}}\right\}
\label{eq:2.6}
\end{eqnarray}
and for longitudinal photons
\begin{eqnarray}
&& \left|\langle \gamma^*|z,{\bm{p}}\rangle - \langle
\gamma^*|z,{\bm{p}}-{\mbox{\boldmath{$\kappa$}}} \rangle\right|^2
_{\lambda_{\gamma}=0} = \nonumber\\
&&8N_c e_f^2\alpha_{em} Q^2z^2(1-z)^2 \left({1   \over
{\bm{p}}^{2}+\varepsilon^{2}}- {1 \over
({\bm{p}}-{\mbox{\boldmath{$\kappa$}}}
)^{2}+\varepsilon^{2}}\right)^2
_{\lambda+\overline{\lambda}=\lambda_{\gamma}}\, , \label{eq:2.7}
\end{eqnarray}
where
\begin{equation} \varepsilon^2 = z(1-z)Q^2 + m_f^2\, .
\label{eq:2.8}
\end{equation}
The leading $\log Q^2$ contribution to the dijet cross section
comes from ${\bm{\Delta}}^2 \alt {\bm{p}}_+^2 +\varepsilon^2$. A
useful small-${\bm{\Delta}}$ expansion for excitation of hard,
${\bm{p}}_+^2 \gg \varepsilon^2=z(1-z)Q^2$, light flavor dijets
from transverse photons is
\begin{eqnarray}
{d\sigma_N \over dz
d^2{\bm{p}}_+ d^2{\bm{\Delta}}} & \approx & {1\over \pi} e_f^2
\alpha_{em}\alpha_S({\bm{p}}_+^2)\left[z^2 + (1-z)^2\right]
\nonumber\\
& \times & {1\over \Delta^4} \cdot{\partial
G(x,{\bm{\Delta}}^2)\over
\partial \log {\bm{\Delta}}^2}\cdot {{\bm{\Delta}}^2 \over (\varepsilon^2
+{\bm{p}}_+^2)^2} \, . \label{eq:2.9}
\end{eqnarray}
Then the single-jet cross section is proportional to the
logarithmic integral
\begin{equation}
{1\over \pi} \int_{0}^{\pi}
d\phi \int^{{\bm{p}}_+^2} {d{\bm{\Delta}}^2 \over {\bm{\Delta}}^2}
\cdot {\partial G(x,{\bm{\Delta}}^2)\over
\partial \log {\bm{\Delta}}^2} = G(x,{\bm{p}}_+^2) \label{eq:2.10}
\end{equation}
familiar from the conventional collinear approximation
\cite{Textbook}.

The small-$x$ result (\ref{eq:2.9}) shows that for a pointlike
projectile of which the photon is just a representative, the
dijets acquire their large transverse momentum from the intrinsic
momentum of the quark and antiquark in the WF of the projectile,
hence dubbing this process a breakup of the photon into hard
dijets is appropriate. The perturbative hard scale $Q_h^2$ for our
process is set by $Q_h^2 \simeq (4{\bm{p}}_+^2+Q^2)$ and
unintegrated gluon density of the proton enters (\ref{eq:2.8}) at
the Bjorken variable $x=(4{\bm{p}}_+^2+Q^2)/W^2$, where $W$ is the
$\gamma^*p$ center of mass energy. On of the major findings of
\cite{Nonlinear} is that the azimuthal decorrelation of dijets
exhibits only a marginal dependence on $Q^2$, and the above
presented formalism is fully applicable to real photons.

Similar formulas apply as well to non-pointlike pions. Indeed, as
argued in \cite{NSSdijet}, the final state interaction between the
final state quark and antiquark can be neglected and the
$q\bar{q}$ plane-wave approximation becomes applicable as soon as
the invariant mass of the forward hard jets exceeds a typical mass
scale of prominent meson resonances. As shown in \cite{NSSdijet}
in the coherent diffractive breakup of pions into hard dijets at
small $x$ the diffractive amplitude is dominated by the
pomeron-splitting mechanism of fig.~1f, when the quark and
antiquark with small intrinsic transverse momentum in the pion
simultaneously acquire large back-to-back transverse momentum from
exchanged gluons \cite{NZsplit} (for the confirmation of the
dominance of the pomeron-splitting mechanism to the higher orders
in pQCD see \cite{Chernyak,Regensburg}). The transverse momentum
distribution in truly inelastic $\pi N$ collisions is different.
In contrast to the pointlike photons, for pions the $q\bar{q}$ WF
$\langle {\bm{p}}| \pi \rangle$ is a soft function which decreases
steeply at ${\bm{p}}^2 > 1/R_{\pi}^2$ (here $R_\pi$ is the pion
radius, for a review of the dominance of the soft WF and
references see \cite{Kroll,Radyushkin}). We are interested in jets
with the transverse momentum much larger than the intrinsic
transverse momentum of (anti)quarks in a pion. The unitarity cuts
of diagrams of fig.~1a-1d show that to the leading order in pQCD
only one parton of the pion  - let it be the quark - can pick up
the large transverse momentum from the exchanged gluon and give
rise to a hard jet in the pion fragmentation region of
pion-nucleon interactions, the spectator jet  retains the soft
intrinsic transverse momentum the antiquark had inside the pion.
Specifically, if the quark jet has a large transverse momentum
${\bm{p}}_+$, then $\langle \pi|z,{\bm{p}}_+\rangle$ can be
neglected, ${\bm{\Delta}} \approx {\bm{p}}_+$ and the pion breakup
cross section takes the form
\begin{eqnarray}
{d\sigma_{\pi N}
\over dz d^2{\bm{p}}_+ d^2{\bm{p}}_-} = { \alpha_S({\bm{p}}_+^2)
\over 2\pi} \cdot {{\cal F}(x,{\bm{p}}_+^2) \over p_+^4}\cdot
\left|\langle \pi |z,{\bm{p}}_-\rangle\right|^2\, .
\label{eq:2.11}
\end{eqnarray}
It shows clearly how the spectator antiquark retains a small
intrinsic transverse momentum it had in the incident pion, as an
analogy cf. the electrodesintegration of the deuteron
\cite{Frullani}. Evidently, excitation of two forward hard jets in
$\pi N$ collisions is only possible to higher orders in pQCD. In
hard inelastic $\pi A$ collisions the higher order pQCD
contributions from multiple scatterings are enhanced by the size
of the extended nuclear target and the purpose of this
communication is a description of the breakup of pions into hard
dijets in inelastic collisions off heavy nuclei within our
nonlinear nuclear $k_{\perp}$-factorization formalism
\cite{Nonlinear}.

The minor technical difference from DIS is the change from the
pointlike $\gamma^{*}q\bar{q}$ vertex $eA_{\mu}\overline{\Psi}
\gamma_{\mu}\Psi$ to the non-pointlike $\pi q\bar{q}$ vertex
$i\Gamma_{\pi}(M^2)\overline{\Psi}\gamma_{5}\Psi$. In terms of the
quark and antiquark helicities $\lambda,\overline{\lambda}$ the
$\pi q({\bm{k}})\bar{q}(-{\bm{k}})$ vertex has the form
(\cite{NSSdijet,Jaus})
\begin{equation}
\overline{\Psi}_{\lambda}({\bm{k}})\gamma_{5}\Psi_{\bar{\lambda}}(-{\bm{k}})
= {\lambda \over \sqrt{z(1-z)}} [m_{f} \delta_{\lambda
-\bar{\lambda}} - \sqrt{2}{\bm{k}}\cdot {\bm{e}}_{-\lambda}
\delta_{\lambda \bar{\lambda}}]\, , \label{eq:2.12}
\end{equation}
where $m_{f}$ is the quark mass and ${\bm{e}}_{\lambda}={1\over
\sqrt{2}} (\lambda {\bm{e}}_{x}+i{\bm{e}}_{y})$ is the familiar
polarization vector for the state of helicity $\lambda$. In
transitions of spin-zero pions into $q\bar{q}$ states with the sum
of helicities $\lambda+\bar{\lambda}=\pm 1$ the latter is
compensated by the orbital momentum of quark and antiquark. In
what follows we shall only need the leading twist term, $\propto
\delta_{\lambda -\bar{\lambda}}$, in (\ref{eq:2.12}), cf. with the
coherent diffractive breakup \cite{NSSdijet}. The corresponding
radial WF $\Psi_{\pi}(z,{\bm{r}})$ is related to the $\pi \to \mu
\nu$ decay constant $F_{\pi}$ and the so-called $\pi$DA
$\varphi_{\pi}(z)$ by
\begin{equation}
\Psi_{\pi}(z,{\bm{r}}=0) = \int {d^2{\bm{p}} \over (2\pi)^2}
\langle z,{\bm{p}} |\pi \rangle = \sqrt{{\pi \over
2N_c}}F_{\pi}\varphi_{\pi}(z)\, . \label{eq:2.13}
\end{equation}
For the purposes of our discussion a convenient normalization is
$\displaystyle \int_0^1 dz \, \varphi_{\pi}(z)=1$. We follow the
Particle Data Group convention $F_{\pi}=131$ MeV \cite{PDG}. The
$\pi$DA depends \cite{PionDA} on the hard scale not shown in
(\ref{eq:2.13}), we shall comment on the relevant scale whenever
appropriate.

\section{The color-dipole $S$-matrix treatment
of the breakup into dijets on nuclear targets}

 We focus on the breakup into dijets at small $x$,
$x\alt x_A = 1/R_A m_N \ll 1$ ($R_A$ is the radius of the nucleus,
$m_N$ is the mass of a nucleon), when the propagation of the
$q\bar{q}$ pair inside nucleus can be treated in the straight-path
approximation. First we review the simpler case of DIS
\cite{Nonlinear}. We work in the conventional approximation of two
t-channel gluons for DIS off free nucleons. The relevant unitarity
cuts of the forward Compton scattering amplitude are shown in
figs. 1a-1d and describe the transition from the color-neutral
$q\bar{q}$ dipole to the color-octet $q\bar{q}$ pair \footnote{To
be more precise, for arbitrary $N_c$ color-excited $q\bar{q}$ pair
is in the adjoint representation and quarks in fundamental
representation of $SU(N_c)$, our reference to the color octet and
triplet must not cause any confusion.}. The unitarity cuts of the
nuclear Compton scattering amplitude which correspond to the
genuine inelastic DIS with color excitation of the nucleus are
shown in figs. 1j,k. The diagram 1k describes multiple color
excitations of a nucleus when the propagating color-octet
$q\bar{q}$ pair rotates in the color space.

Let ${\bm{b}}_+$ and ${\bm{b}}_-$ be the impact parameters of the
quark and antiquark, respectively, and
$S_A(\{{\bm{b}}_j\},{\bm{b}}_+,{\bm{b}}_-)$  be the S-matrix for
interaction of the $q\bar{q}$ pair with the nucleus where
$\{{\bm{b}}_j\}$ stands for the positions of nucleons. The initial
state $|A;1\rangle$ is a color-singlet nucleus made of
color-singlet nucleons and a color-singlet $q\bar{q}$ dipole, in
the final state we sum over all excitations of the target nucleus
when one or several nucleons have been color excited. A convenient
way to sum such cross sections is offered by the closure relation
\cite{Glauber}. Regarding the color states $c_{km}$ of the
$q_k\bar{q}_m$ pair, we sum over all octet and singlet states.
Then, the  2-body inclusive spectrum is calculated in terms of the
2-body density matrix as
\begin{eqnarray}
&&{d\sigma_{inel} \over dz d^2{\bm{p}}_+ d^2{\bm{p}}_-} = {1\over
(2\pi)^4} \int d^2 {\bm{b}}_+' d^2{\bm{b}}_-' d^2{\bm{b}}_+
d^2{\bm{b}}_-   \exp[-i{\bm{p}}_+({\bm{b}}_+
-{\bm{b}}_+')-i{\bm{p}}_-({\bm{b}}_- - {\bm{b}}_-')]\nonumber \\
&& \times \Psi^*(z,{\bm{b}}_+' -{\bm{b}}_-') \Psi(z,{\bm{b}}_+
-{\bm{b}}_-)\,
\Omega^{inel}(\bm{b}_+',\bm{b}_-',\bm{b}_+,\bm{b}_-) \, ,
\label{eq:3.1}
\end{eqnarray}
where the superscript $inel$ refers to the truly inelastic cross
section, with the contribution from diffractive processes
subtracted. The projectile wave function $\Psi$ in general carries
a dependence on helicities, flavor, and for the photon, virtuality
$Q^2$, which have not been put in evidence here.
The generalized cross section operator
$\Omega^{inel}(\bm{b}_+',\bm{b}_-',\bm{b}_+,\bm{b}_-)$  is
expressed through the $q\bar{q}$--nucleus $S$--matrix as
\begin{eqnarray}
&& \Omega^{inel}(\bm{b}_+',\bm{b}_-',\bm{b}_+,\bm{b}_-) =
\sum_{A^*} \sum_{km}\langle
1;A|S_A^*(\{{\bm{b}}_j\},{\bm{b}}_+',{\bm{b}}_-')|A^*;c_{km}\rangle
\langle
c_{km};A^*|S_A(\{{\bm{b}}_j\},{\bm{b}}_+,{\bm{b}}_-)|A;1\rangle  \nonumber\\
&& -  \langle 1;A|S_A^*(\{{\bm{b}}_j\},{\bm{b}}_+',{\bm{b}}_-')
|A;1\rangle \langle
1;A|S_A(\{{\bm{b}}_j\},{\bm{b}}_+,{\bm{b}}_-)|A;1\rangle
\label{eq:3.1a}
\end{eqnarray}
Upon the application of closure in the sum over nuclear states,
the first term in eq.(\ref{eq:3.1a}) becomes
\begin{eqnarray}
&&\sum_{A^*} \sum_{km} \langle A| \left\{ \langle 1|
S_A^*(\{{\bm{b}}_j\},{\bm{b}}_+',{\bm{b}}_-')|c_{km}\rangle \}
|A^* \rangle \langle A^*| \{\langle c_{km}|
S_A(\{{\bm{b}}_j\},{\bm{b}}_+,{\bm{b}}_-)|1\rangle \right\}
|A\rangle =\nonumber\\
&&=\langle A| \left\{ \sum_{km} \langle 1|
S_A^*(\{{\bm{b}}_j\},{\bm{b}}_+',{\bm{b}}_-')|c_{km}\rangle
\langle
c_{km}|S_A(\{{\bm{b}}_j\},{\bm{b}}_+,{\bm{b}}_-)|1\rangle\right\}
|A\rangle \, ,
\label{eq:3.2}
\end{eqnarray}
and can be considered as an intranuclear evolution operator for
the 2-body density matrix.

The further analysis of (\ref{eq:3.2}) is a non-Abelian
generalization of the formalism developed by one of the authors
(BGZ) for the in-medium evolution of ultrarelativistic positronium
\cite{BGZpositronium}. Let the QCD eikonal for the quark-nucleon
and antiquark-nucleon one-gluon exchange interaction be $T^a_{+}
\Delta({\bm{b}})$ and  $T^a_{-} \Delta({\bm{b}})$, where $T^{a}_+$
and $T^{a}_{-}$ are the $SU(N_c)$ generators for the quark and
antiquarks states, respectively. The vertex $V_a$ for excitation
of the nucleon $g^a N \to N^*_a$ into a color octet state is so
normalized that after application of closure the vertex $g^a g^b
NN$ in the diagrams of fig. 1a-d is $\delta_{ab}$. Then, to the
two-gluon exchange approximation, the $S$-matrix of the
$(q\bar{q})$-nucleon interaction equals
\begin{eqnarray}
S_N({\bm{b}}_+,{\bm{b}}_-) = 1 + i[T^a_{+} \Delta({\bm{b}}_+)+
T^a_{-} \Delta({\bm{b}}_-)]V_a - {1\over 2} [T^a_{+}
\Delta({\bm{b}}_+)+ T^a_{-} \Delta({\bm{b}}_-)]^2\, .
\label{eq:3.3}
\end{eqnarray}
The profile function for interaction of the $q\bar{q}$ dipole with
a nucleon is $\Gamma({\bm{b}}_+,{\bm{b}}_-)= 1 -
S_N({\bm{b}}_+,{\bm{b}}_-)$ and the dipole cross section for
color-singlet $q\bar{q}$ dipole equals
\begin{equation}
\sigma({\bm{b}}_+-{\bm{b}}_-) = 2\int
d^2{\bm{b}}_{+} \langle N|\Gamma({\bm{b}}_+,{\bm{b}}_-) |N\rangle
= {N_c^2 -1 \over 2N_c}\int d^2{\bm{b}}_{+}
[\Delta({\bm{b}}_+)-\Delta({\bm{b}}_-)]^2\,. \label{eq:3.4}
\end{equation}
The nuclear $S$-matrix of the straight-path approximation for the
dilute-gas nucleus is given by \cite{Glauber}
\begin{eqnarray}
S_A(\{{\bm{b}}_j\},{\bm{b}}_+,{\bm{b}}_-) = \prod _{j=1}^A
S_N({\bm{b}}_+ -{\bm{b}}_j,{\bm{b}}_- - {\bm{b}}_j) \label{eq:3.5}
\end{eqnarray}
where the ordering along the longitudinal path is understood. To
the two-gluon exchange approximation, only the terms quadratic in
$\Delta({\bm{b}}_j)$ must be kept in the evaluation of the
single-nucleon matrix elements
$$
\langle N_j|S_N^*({\bm{b}}_+' -{\bm{b}}_j,{\bm{b}}_-' -
{\bm{b}}_j) S_N({\bm{b}}_+-{\bm{b}}_j,{\bm{b}}_- - {\bm{b}}_j)|N_j
\rangle
$$
which enter the calculation of $S_A^*S_A$. The evolution operator
for the 2-body density matrix (\ref{eq:3.2}) equals the $S$-matrix
$S_{4A}({\bm{b}}_+,{\bm{b}}_-,{\bm{b}}_+',{\bm{b}}_-')$ for
scattering of a fictitious 4-parton state composed of two
quark-antiquark pairs in an overall color-singlet state
\cite{BGZpositronium,NPZcharm,LPM}. Namely, because $(T^{a}_+)^*=
-T^{a}_{-}$, within the two-gluon exchange approximation the
quarks entering the complex-conjugate $S_A^*$ in (\ref{eq:3.2})
can be viewed as antiquarks, so that
\begin{eqnarray}
\sum_{km} \langle 1|
S_A^*(\{{\bm{b}}_j\},{\bm{b}}_+',{\bm{b}}_-')|c_{km}\rangle
\langle c_{km}|S_A(\{{\bm{b}}_j\},{\bm{b}}_+,{\bm{b}}_-)|1\rangle
=\nonumber\\
=\sum_{km jl} \delta_{kl} \delta_{mj} \langle
c_{km}c_{jl}|S_{4A}({\bm{b}}_+',{\bm{b}}_-',{\bm{b}}_+,{\bm{b}}_-)|11\rangle
\, . \label{eq:3.6}
\end{eqnarray}
While the first $q\bar{q}$ pair is formed by the initial quark $q$
and antiquark $\bar{q}$ at impact parameters ${\bm{b}}_{+}$ and
${\bm{b}}_-$, respectively, in the second pair $q'\bar{q}'$ the
quark $q'$ propagates at an impact parameter ${\bm{b}}_-'$ and the
antiquark $\bar{q}'$ at an impact parameter ${\bm{b}}_+'$. In the
initial state both the $q\bar{q}$ and $q'\bar{q}'$ pairs are in
color-singlet states: $ | in \rangle = |11\rangle$. The sum over
color states of the produced quark-antiquark pair can be
represented as
\begin{eqnarray}
\sum_{km} \langle c_{km}
c_{km}|S_{4A}({\bm{b}}_+',{\bm{b}}_-',{\bm{b}}_+,{\bm{b}}_-)|11\rangle
=
\langle 11|S_{4A}({\bm{b}}_+',{\bm{b}}_-',{\bm{b}}_+,{\bm{b}}_-)|11\rangle \nonumber\\
+ \sqrt{N_c^2 -1} \langle
88|S_{4A}({\bm{b}}_+',{\bm{b}}_-',{\bm{b}}_+,{\bm{b}}_-)|11\rangle
\, . \label{eq:3.7}
\end{eqnarray}
Let $\sigma_4({\bm{b}}_+',{\bm{b}}_-',{\bm{b}}_+,{\bm{b}}_-)$ be
the color-dipole cross section operator for the 4-body state. It
is convenient to introduce the average impact parameter
\begin{equation}
{\bm{b}} = {1\over 4}( {\bm{b}}_+ + {\bm{b}}_+' +
{\bm{b}}_- + {\bm{b}}_-')\, , \label{eq:3.8}
\end{equation}
and
\begin{equation} {\bm{s}} = {\bm{b}}_+ - {\bm{b}}_+' \, ,
\label{eq:3.9}
\end{equation}
for the variable conjugate to the decorrelation momentum, in terms
of which
\begin{eqnarray}
{\bm{b}}_+ - {\bm{b}}_-' = {\bm{s}} + {\bm{r}}'\, ,~~ {\bm{b}}_- - {\bm{b}}_+'
= {\bm{s}} - {\bm{r}} \, ,~~ {\bm{b}}_- - {\bm{b}}_-' = {\bm{s}}
-{\bm{r}} + {\bm{r}}'\, . \label{eq:3.10}
\end{eqnarray}
Then the standard evaluation of the nuclear expectation value for
a dilute gas nucleus neglecting the size of color dipoles compared
to the radius of a heavy nucleus gives \cite{Glauber}
\begin{eqnarray}
S_{4A}({\bm{b}}_+',{\bm{b}}_-',{\bm{b}}_+,{\bm{b}}_-)=\exp[-
{1\over 2}\sigma_{4}({\bm{s}},{\bm{r}},{\bm{r}}')T({\bm{b}})]
\label{eq:3.11}
\end{eqnarray}
where
\begin{eqnarray}
T({\bm{b}})=\int db_z n_{A}(b_z, {\bm{b}})
\label{eq:3.12}
\end{eqnarray}
is the optical thickness of a nucleus at an impact parameter
${\bm{b}}$, the nuclear matter density $n_A(b_z,{\bm{b}})$ is so
normalized that $\int db_z d^2{\bm{b}} n_{A}(z, {\bm{b}}) = A$.
The single-nucleon $S$-matrix (\ref{eq:3.3}) contains transitions
from the color-singlet to the both color-singlet and color-octet
$q\bar{q}$ pairs. However, only the color-singlet operators
contribute to $\langle N_j|S_N^*({\bm{b}}_+'
-{\bm{b}}_j,{\bm{b}}_-' - {\bm{b}}_j)
S_N({\bm{b}}_+-{\bm{b}}_j,{\bm{b}}_- - {\bm{b}}_j)|N_j \rangle$,
and the matrix $\sigma_4({\bm{s}},{\bm{r}},{\bm{r}}')$ only
includes transitions between the $|11\rangle$ and $|88\rangle$
color-singlet 4-parton states.

The calculation of $\sigma_4({\bm{s}},{\bm{r}},{\bm{r}}')$ is
found in \cite{Nonlinear}, here we only cite the results:
\begin{eqnarray}
\sigma_{11}=\langle 11|\sigma_4|11\rangle =
&& \sigma({\bm{r}})+\sigma({\bm{r}}')\, , \label{eq:3.13} \\
\sigma_{88}=\langle 88|\sigma_4|88\rangle  = &&{N_c^2 -2 \over
N_c^2-1} [\sigma({\bm{s}})+ \sigma({\bm{s}} -{\bm{r}} +
{\bm{r}}')] + {2 \over N_c^2-1}[\sigma({\bm{s}} + {\bm{r}}')+
\sigma({\bm{s}} - {\bm{r}})] \nonumber \\
&& - {1\over N_c^2-1}
[\sigma({\bm{r}})+\sigma({\bm{r}}')]\,, \label{eq:3.14} \\
\sigma_{18}=\sigma_{81}= \langle 11|\sigma_4|88\rangle
= &&{1\over \sqrt{N_c^2-1}} \left[\sigma({\bm{s}})-
\sigma({\bm{s}} + {\bm{r}}')- \sigma({\bm{s}} -{\bm{r}}) +
\sigma({\bm{s}} - {\bm{r}} + {\bm{r}}')\right] \nonumber \\
\equiv && -{\Sigma_{18}({\bm{s}},{\bm{r}},{\bm{r}}') \over
\sqrt{N_c^2-1}}\, . \label{eq:3.15}
\end{eqnarray}
The term in (\ref{eq:3.1}), which subtracts the contribution from
processes without color excitation of the target nucleus, equals
\begin{eqnarray}
\langle 1;A|S_A^*({\bm{b}}_+',{\bm{b}}_-') |A;1\rangle \langle
1;A| S_A({\bm{b}}_+,{\bm{b}}_-)|A;1\rangle &&= \exp\{- {1\over
2}\left[\sigma({\bm{r}})+\sigma({\bm{r}}')\right]T({\bm{b}})\}
\nonumber \\
&&= \exp[- {1\over 2}\sigma_{11}T({\bm{b}})] \label{eq:3.16}
\end{eqnarray}
It is convenient to use the Sylvester expansion
\begin{eqnarray}
\exp[- {1\over 2}\sigma_4 T({\bm{b}})] = \exp[- {1\over 2}\Sigma_1
T({\bm{b}})]{\sigma_4 - \Sigma_2 \over \Sigma_1 - \Sigma_2} +
\exp[- {1\over 2}\Sigma_2 T({\bm{b}})]{\sigma_4 - \Sigma_1 \over
\Sigma_2 - \Sigma_1}
\label{eq:3.17}
\end{eqnarray}
where $\Sigma_{1,2}$ are the two eigenvalues of
the operator $\sigma_4$,
\begin{equation}
\Sigma_{1,2} = {1\over 2} (\sigma_{11}+\sigma_{88}) \mp {1\over
2}(\sigma_{11}-\sigma_{88}) \sqrt{1 + {4\sigma_{18}^2 \over
(\sigma_{11}-\sigma_{88})^2}}\, .
\label{eq:3.18}
\end{equation}
An application to (\ref{eq:3.7}) of the Sylvester expansion gives
for the function $\Omega^{inel}$ in the integrand of
(\ref{eq:3.1})
\begin{eqnarray}
\Omega^{inel}(\bm{b}_+',\bm{b}_-',\bm{b}_+,\bm{b}_-) = && (\langle
11| + \sqrt{N_c^2 -1} \langle 88|) \exp\left[-{1\over 2}\sigma_4
T({\bm{b}})\right]|11\rangle -
\exp\left[-{1\over 2}\sigma_{11}T({\bm{b}})\right]\nonumber\\
= && \exp\left[-{1\over 2}\Sigma_2 T({\bm{b}})\right]-
\exp\left[-{1\over 2}\sigma_{11}T({\bm{b}})\right]\nonumber\\
&&+ { \sigma_{11} - \Sigma_2 \over \Sigma_1 - \Sigma_2}
\left\{\exp\left[-{1\over 2}\Sigma_1 T({\bm{b}})\right]-
\exp\left[-{1\over 2}\Sigma_2 T({\bm{b}})\right]\right\}\nonumber\\
&&+{\sqrt{N_c^2 -1}  \sigma_{18} \over \Sigma_1 - \Sigma_2}
\left\{\exp\left[-{1\over 2}\Sigma_1 T({\bm{b}})\right]-
\exp\left[-{1\over 2}\Sigma_2 T({\bm{b}})\right]\right\}
\label{eq:3.19}
\end{eqnarray}
Notice that the difference between $\Sigma_2$ and
$\sigma_{11}=\sigma({\bm{r}})+\sigma({\bm{r}}')$ is quadratic or
higher order in the off-diagonal $\sigma_{18}$, see
eq.~(\ref{eq:3.18}). Consequently, the first two lines in the
Sylvester expansion (\ref{eq:3.19}) start with terms $\propto
\sigma_{18}^2$, whereas the last line starts with terms $\propto
\sigma_{18}$. Then it is convenient to represent (\ref{eq:3.19})
as an impulse approximation (IA) term times a nuclear distortion
factor $D_{A}({\bm{s}},{\bm{r}},{\bm{r}}',{\bm{b}})$,
\begin{eqnarray}
\Omega^{inel}(\bm{b}_+',\bm{b}_-',\bm{b}_+,\bm{b}_-)\equiv
\Sigma_{18}({\bm{s}},{\bm{r}},{\bm{r}}')
D_{A}({\bm{s}},{\bm{r}},{\bm{r}}',{\bm{b}})\, , \label{eq:3.20}
\end{eqnarray}
so that
\begin{eqnarray}
{d\sigma^{inel} \over d^2{\bm{b}} dz d^2{\bm{p}}_+ d^2{\bm{p}}_-}
&=& {1\over 2(2\pi)^4} \int d^2{\bm{s}} d^2{\bm{r}}
d^2{\bm{r}}'\exp[-i({\bm{p}}_+ +{\bm{p}}_-){\bm{s}}
+i{\bm{p}}_-({\bm{r}}' -{\bm{r}})]
\nonumber\\
&\times& \Psi^*(z,{\bm{r}}') \Psi(z,{\bm{r}})
T({\bm{b}})D_A({\bm{s}},{\bm{r}},{\bm{r}}',{\bm{b}})\Sigma_{18}({\bm{s}},{\bm{r}},{\bm{r}}')
\, . \label{eq:3.21}
\end{eqnarray}
What we need is a Fourier representation for each and every factor
in (\ref{eq:3.21}).


\section{Nonlinear
$k_{\perp}$-factorization for breakup into dijets and collective
WW glue of nuclei}

Upon the application of  (\ref{eq:2.1}) the IA factor in
(\ref{eq:3.3}) admits the simple Fourier representation
\begin{eqnarray}
\Sigma_{18}({\bm{s}},{\bm{r}},{\bm{r}}')=\sigma({\bm{s}} + {\bm{r}}')+
\sigma({\bm{s}} -{\bm{r}})- \sigma({\bm{s}})- \sigma({\bm{s}} -
{\bm{r}} + {\bm{r}}')
\nonumber\\
= \alpha_S \sigma_{0}(x) \int d^2{\bm{\kappa}}
f({\mbox{\boldmath{$\kappa$}}})
\exp[i{\mbox{\boldmath{$\kappa$}}}{\bm{s}}] \left\{1 -
\exp[i{\mbox{\boldmath{$\kappa$}}} {\bm{r}}']\right\} \left\{1 -
\exp[-i{\mbox{\boldmath{$\kappa$}}} {\bm{r}}]\right\}\, ,
\label{eq:4.1}
\end{eqnarray}
The  Fourier representation of the nuclear distortion factor in
terms of the collective nuclear WW gluon distribution as defined
in \cite{NSSdijet,Saturation} is not a trivial task, though,
appealing analytic results are derived in the large-$N_c$
approximation.

The crucial point is that in the large-$N_c$ approximation
$\Sigma_1 = \sigma({\bm{s}})+\sigma({\bm{s}}+{\bm{r}}'-{\bm{r}})$
and $\Sigma_{2} = \Sigma_{22} =
\sigma({\bm{r}})+\sigma({\bm{r}}')$, so that only the last term in
the Sylvester expansion (\ref{eq:3.19}) contributes to the jet-jet
inclusive cross section. At large $N_c$ the initial color singlet
dipole excites to the color--octet state and further intranuclear
color exchanges only rotate the dipole between different
color--octet states. This is indicated schematically in fig. 2.
Then the nuclear distortion factor takes on a simple form
\begin{eqnarray}
D_A({\bm{s}},{\bm{r}},{\bm{r}}',{\bm{b}}) = { 2 \over (\Sigma_2 -
\Sigma_1)T({\bm{b}})} \left\{\exp\left[-{1\over 2}\Sigma_1
T({\bm{b}})\right]- \exp\left[-{1\over 2}\Sigma_2
T({\bm{b}})\right]\right\} \, .
\label{eq:4.2}
\end{eqnarray}
The denominator $(\Sigma_2 - \Sigma_1)$ is problematic from the
point of view of the Fourier transform, but it can be eliminated
by the integral representation
\begin{eqnarray}
D_A({\bm{s}},{\bm{r}},{\bm{r}}',{\bm{b}}) &=& \int_0^1 d \beta
\exp\left\{-{1\over 2}[\beta \Sigma_1
+(1-\beta)\Sigma_2]T({\bm{b}})\right\} =\nonumber\\
&=& \int_0^1 d \beta \exp\left\{-{1\over
2}(1-\beta)[\sigma({\bm{r}})+\sigma({\bm{r}}')]
T({\bm{b}})\right\}  \nonumber\\
&\times&\exp\left\{-{1\over
2}\beta[\sigma({\bm{s}})+\sigma({\bm{s}}+{\bm{r}}'-{\bm{r}})]
T({\bm{b}})\right\} \, .
\label{eq:4.3}
\end{eqnarray}
Here the former two exponential factors describe the initial state
(IS) intranuclear distortion of the incoming color-singlet
$(q\bar{q})$ dipole state, whereas the last two factors describe
the FS distortion of the outgoing color-octet states.

Next we apply to the exponential factors in (\ref{eq:4.3}) the NSS
representation in terms of the collective WW unintegrated gluon
density of the nucleus \cite{Saturation,NSSdijet}:
\begin{eqnarray}
\exp\left[-{1\over 2}\sigma({\bm{s}}) T({\bm{b}})\right] = \int
d^2{\mbox{\boldmath{$\kappa$}}} \Phi(\nu_{A}({\bm{b}}),
{\mbox{\boldmath{$\kappa$}}})
\exp(i{\mbox{\boldmath{$\kappa$}}}{\bm{s}})  \, ,
\label{eq:4.4}
\end{eqnarray}
where
\begin{eqnarray}
\Phi(\nu_{A}({\bm{b}}), {\mbox{\boldmath{$\kappa$}}})= \sum_{j
\geq 0}
w_{j}(\nu_A({\bm{b}}))f^{(j)}({\mbox{\boldmath{$\kappa$}}}) =
\exp(-\nu_{A}({\bm{b}}))f^{(0)}({\mbox{\boldmath{$\kappa$}}})
+\phi_{WW}(\nu_{A}({\bm{b}}),{\mbox{\boldmath{$\kappa$}}})\, .
\label{eq:4.5}
\end{eqnarray}
Here $\phi_{WW}(\nu_{A}({\bm{b}}),{\mbox{\boldmath{$\kappa$}}})$
is the unintegrated collective nuclear Weizs\"acker-Williams glue
per unit area in the impact parameter plane,
\begin{eqnarray}
w_{j}(\nu_A({\bm{b}}))= {\nu_{A}^{j}({\bm{b}}) \over
j!}\exp\left[-\nu_{A}({\bm{b}})\right]
\label{eq:4.6}
\end{eqnarray}
is a probability of finding $j$ spatially overlapping nucleons at
an impact parameter ${\bm{b}}$ in a Lorentz-contracted nucleus,
\begin{eqnarray}
\nu_{A}({\bm{b}})= {1\over 2}\alpha_S(r)\sigma_0(x) T({\bm{b}}) \,
, \label{eq:4.7}
\end{eqnarray}
and
\begin{equation}
f^{(j)}({\mbox{\boldmath{$\kappa$}}} )= \int \prod_{i=1}^j
d^2{\mbox{\boldmath{$\kappa$}}} _{i}
f({\mbox{\boldmath{$\kappa$}}} _{i})
\delta({\mbox{\boldmath{$\kappa$}}} -\sum_{i=1}^j
{\mbox{\boldmath{$\kappa$}}} _i) \,,
~~f^{(0)}({\mbox{\boldmath{$\kappa$}}})=\delta({\mbox{\boldmath{$\kappa$}}})
\label{eq:4.8}
\end{equation}
is a collective gluon field of $j$ overlapping nucleons. As shown
in \cite{NSSdijet,Saturation} the collective nuclear unintegrated
gluon density
$\phi_{WW}(\nu_{A}({\bm{b}}),{\mbox{\boldmath{$\kappa$}}})$ enters
the calculation of the nuclear sea quark density in precisely the
same way as $f({\mbox{\boldmath{$\kappa$}}})$ in (\ref{eq:2.5})
for the free nucleon target.

We cite two important features of
$\phi_{WW}(\nu_{A}({\bm{b}}),{\mbox{\boldmath{$\kappa$}}})$.
First, the hard tail of the unintegrated nuclear glue per bound
nucleon is calculated parameter free \cite{NSSdijet},
\begin{eqnarray}
f_{WW}(\nu_A({\bm{b}}),{\mbox{\boldmath{$\kappa$}}} ) =
{\phi_{WW}(\nu_A({\bm{b}}),{\mbox{\boldmath{$\kappa$}}}) \over
\nu_A({\bm{b}})} = f({\mbox{\boldmath{$\kappa$}}} )\left[1+ {2
C_A\pi^2\gamma^2\alpha_S(r)T({\bm{b}})\over C_F N_c
{\mbox{\boldmath{$\kappa$}}}^2}
G({\mbox{\boldmath{$\kappa$}}}^2)\right] \,,
\label{eq:4.9}
\end{eqnarray}
and does not depend on the infrared parameter $\sigma_{0}(x)$. In
the hard regime the differential nuclear glue is not shadowed,
furthermore, because of the manifestly positive-valued and
model-independent nuclear higher twist correction it exhibits a
nuclear antishadowing property \cite{NSSdijet}. Second, for
interactions with nuclei of $q\bar{q}$ dipoles with $|{\bm{r}}|
\agt 1/Q_A$, the strong coupling enters (\ref{eq:4.7}) as
$\alpha_S(Q_A^2)$ and in the saturation region of
${\mbox{\boldmath{$\kappa$}}}^2 \alt Q_A^2$ we have
\cite{Saturation,Nonlinear}
\begin{eqnarray}
\Phi(\nu_{A}({\bm{b}}), {\mbox{\boldmath{$\kappa$}}}) \approx
\phi_{WW}(\nu_{A}({\bm{b}}),{\mbox{\boldmath{$\kappa$}}}) \approx
{1\over \pi}{Q_{A}^2 \over
({\mbox{\boldmath{$\kappa$}}}^2+Q_{A}^2)^2}\, ,
\label{eq:4.10}
\end{eqnarray}
where the width of the plateau $Q_A$ which is the nuclear
saturation scale equals
\begin{eqnarray}
Q_A^2 \approx { 4\pi^2 \over N_c} \alpha_S(Q_A^2)G(Q_A^2)
T({\bm{b}})
\label{eq:4.11}
\end{eqnarray}
and exhibits only weak dependence on the infrared parameters
through the $Q_A^2$ dependence of the running strong coupling and
scaling violations in the unintegrated gluon density of the
nucleon. For instance, at $x=10^{-2}$ the numerical results
\cite{INDiffGlue} for $G(Q^2)$ correspond to a nearly $Q^2$
independent
 $ \alpha_S(Q^2)G(Q^2)\approx 1$. For average DIS on a heavy nucleus,
$A^{1/3}=6$, we found $\langle Q_A^2({\bm{b}})\rangle \approx 0.9$
$(GeV/c)^2$.

Now we are in the position to represent the nuclear distortion
factor (\ref{eq:4.3}) as
\begin{eqnarray}
D_A({\bm{s}},{\bm{r}},{\bm{r}}',{\bm{b}})= \int_0^1 d \beta &&\int
d^2{\mbox{\boldmath{$\kappa$}}}_1
\Phi((1-\beta)\nu_A({\bm{b}}),{\mbox{\boldmath{$\kappa$}}}_1)
\exp(-i{\mbox{\boldmath{$\kappa$}}}_1 {\bm{r}})
\nonumber\\
\times &&\int d^2{\mbox{\boldmath{$\kappa$}}}_2
\Phi((1-\beta)\nu_A({\bm{b}}),{\mbox{\boldmath{$\kappa$}}}_2)
\exp(i{\mbox{\boldmath{$\kappa$}}}_2 {\bm{r}}')
\nonumber\\
\times &&\int d^2{\mbox{\boldmath{$\kappa$}}}_3
\Phi(\beta\nu_A({\bm{b}}),{\mbox{\boldmath{$\kappa$}}}_3)
\exp[i{\mbox{\boldmath{$\kappa$}}}_3({\bm{s}}+{\bm{r}}'-
{\bm{r}})]
\nonumber\\
\times &&\int d^2{\mbox{\boldmath{$\kappa$}}}_4
\Phi(\beta\nu_A({\bm{b}}),{\mbox{\boldmath{$\kappa$}}}_4)
\exp(i{\mbox{\boldmath{$\kappa$}}}_4 {\bm{s}}) \, ,
\label{eq:4.12}
\end{eqnarray}
so that the jet-jet inclusive inelastic cross section takes the
form
\begin{eqnarray}
&&{d\sigma_{inel} \over d^2{\bm{b}} dz d^2{\bm{p}}_{-}
d^2{\bm{\Delta}}} = {1\over 2(2\pi)^2} \alpha_S \sigma_0(x)
T({\bm{b}})\int_0^1 d \beta \int d^2{\mbox{\boldmath{$\kappa$}}}_3
d^2{\mbox{\boldmath{$\kappa$}}} f({\mbox{\boldmath{$\kappa$}}})
\nonumber\\
&&\times \, \, \,  \Phi(\beta\nu_A({\bm{b}}),{\bm{\Delta}}
-{\mbox{\boldmath{$\kappa$}}}_3 -{\mbox{\boldmath{$\kappa$}}})
\Phi(\beta\nu_A({\bm{b}}),{\mbox{\boldmath{$\kappa$}}}_3)
\nonumber \\
&& \times\Biggr|\int d^2{\mbox{\boldmath{$\kappa$}}}_1
\Phi((1-\beta)\nu_A({\bm{b}}),{\mbox{\boldmath{$\kappa$}}}_1)
\left\{\langle \gamma^*| z,{\bm{p}}_{-}
+{\mbox{\boldmath{$\kappa$}}}_1 +{\mbox{\boldmath{$\kappa$}}}_3
\rangle - \langle \gamma^*| z,{\bm{p}}_{-}
+{\mbox{\boldmath{$\kappa$}}}_1
+{\mbox{\boldmath{$\kappa$}}}_3+{\mbox{\boldmath{$\kappa$}}}
\rangle \right\}\Biggl|^2\, . \label{eq:4.13}
\end{eqnarray}
The transverse momentum distribution of dijets is uniquely
calculable in terms of the collective WW glue of a nucleus and as
such (\ref{eq:4.13}) can be regarded as a nonlinear nuclear
$k_{\perp}$-factorization for the inclusive inelastic dijet cross
section. Notice that the convolution in the last line of eq.
(\ref{eq:4.13}) describes the initial state distortion of the
color singlet $q\bar{q}$--state in the projectile.

There are two important limiting cases. We start with  hard
dijets, $|{\bm{p}}_{\pm}| \agt Q_A$. A crucial point is that the
WF of the pointlike photon is a slowly decreasing function of the
transverse momentum, in contrast to
$\Phi(\nu_A({\bm{b}}),{\mbox{\boldmath{$\kappa$}}})$ which is a
steeply decreasing function, compare
eqs.~(\ref{eq:2.6}),(\ref{eq:2.7}) to eq.~(\ref{eq:4.10}). Then,
since ${\mbox{\boldmath{$\kappa$}}}_i^2 \alt Q_A^2$, for hard
dijets one can neglect
${\mbox{\boldmath{$\kappa$}}}_1,{\mbox{\boldmath{$\kappa$}}}_3$
compared to ${\bm{p}}_{\pm}$ and approximate
\begin{eqnarray}
\int d^2{\mbox{\boldmath{$\kappa$}}}_1
\Phi((1-\beta)\nu_A({\bm{b}}),{\mbox{\boldmath{$\kappa$}}}_1)
\left\{\langle \gamma^*| z,{\bm{p}}_{-}
+{\mbox{\boldmath{$\kappa$}}}_1 +{\mbox{\boldmath{$\kappa$}}}_3
\rangle - \langle \gamma^*| z,{\bm{p}}_{-}
+{\mbox{\boldmath{$\kappa$}}}_1
+{\mbox{\boldmath{$\kappa$}}}_3+{\mbox{\boldmath{$\kappa$}}}
\rangle
\right\} \nonumber\\
\approx \langle \gamma^*| z,{\bm{p}}_{-}\rangle - \langle
\gamma^*| z,{\bm{p}}_{-} +{\mbox{\boldmath{$\kappa$}}} \rangle \,
, \, .
\label{eq:4.14}
\end{eqnarray}
which amounts to negligible IS distortion of small color-singlet
dipoles with $|{\bm{r}}|,|{\bm{r}}'| \sim 1/|{\bm{p}}_\pm| \alt
1/Q_A$. Next we notice that
\begin{equation}
\int d^2{\mbox{\boldmath{$\kappa$}}}_3\,
\Phi(\beta\nu_A({\bm{b}}),{\bm{\Delta}}
-{\mbox{\boldmath{$\kappa$}}}_3 -{\mbox{\boldmath{$\kappa$}}})
\Phi(\beta\nu_A({\bm{b}}),{\mbox{\boldmath{$\kappa$}}}_3) =
\Phi(2\beta\nu_A({\bm{b}}),{\bm{\Delta}}
-{\mbox{\boldmath{$\kappa$}}})\, ,
\label{eq:4.15}
\end{equation}
so that the hard jet-jet inclusive cross section takes the form
\begin{eqnarray}
{d\sigma_{inel} \over d^2{\bm{b}} dz d^2{\bm{p}}_+
d^2{\bm{\Delta}}}= T({\bm{b}}) \int_0^1 d \beta \int
d^2{\mbox{\boldmath{$\kappa$}}} \, \Phi(2\beta
\nu_A({\bm{b}}),{\bm{\Delta}} - {\mbox{\boldmath{$\kappa$}}})
{d\sigma_{N} \over dz d^2{\bm{p}}_+
d^2{\mbox{\boldmath{$\kappa$}}} } \, , \label{eq:4.16}
\end{eqnarray}
which is a close counterpart of, but still different from, the
conventional $k_{\perp}$-factorization (\ref{eq:2.3}) for the free
nucleon target. As a matter of fact, for hard dijets one does not
need to invoke the large--$N_c$ approximation: here
$|\bm{r}|,|\bm{r}'| \ll |\bm{s}|$, so that $\Sigma_1 \sim 0$,
$\Sigma_2 \sim 2 \lambda_c \sigma(\bm{s})$, where $\lambda_c
=N_c^2/(N_c^2-1)$, and one can replace $ \Phi(2\beta
\nu_A({\bm{b}}),{\bm{\Delta}} - {\mbox{\boldmath{$\kappa$}}})$ by
$ \Phi(2\lambda_c \beta  \nu_A({\bm{b}}),{\bm{\Delta}} -
{\mbox{\boldmath{$\kappa$}}})$, see a discussion in
\cite{Nonlinear}. All the dependence on transverse momentum $p_+$
of the hard jet is in the free nucleon cross section $d\sigma_N$,
i.e., the  pQCD treatment breakup into hard dijets is applicable
to DIS on the free nucleon and nuclear targets on the same
footing. The effect of the collective nuclear glue $\Phi(2\beta
\nu_A({\bm{b}}),{\mbox{\boldmath{$\kappa$}}})$ is a
smearing/broadening as well as decorrelation of the dijets.
Numerical estimates for the azimuthal decorrelation of jets and a
discussion concerning the relevance to the RHIC-STAR finding
\cite{RHIC_STAR} of the disappearance of the away jet in central
$AuAu$ collisions can be found in  \cite{Nonlinear}. In this hard
dijet limit it is tempting to assign to $\Phi(2\beta
\nu_A({\bm{b}}),{\mbox{\boldmath{$\kappa$}}})$ a probabilistic
interpretation as an intrinsic transverse momentum distribution of
collective nuclear gluons, but this collectivization only applies
to the fraction $\beta$ of the nuclear thickness which the
$(q\bar{q})$ pair propagates in the color octet state.

The second limiting case is that of minijets
$|{\bm{p}}_-|,|{\bm{\Delta}}| \alt Q_A$. Since
$|{\mbox{\boldmath{$\kappa$}}}_{i}|\sim Q_A$ one can neglect
${\bm{p}}_-$ in the photon's wave functions,
\begin{eqnarray}
\int d^2{\mbox{\boldmath{$\kappa$}}}_1
\Phi((1-\beta)\nu_A({\bm{b}}),{\mbox{\boldmath{$\kappa$}}}_1)
\left\{\langle \gamma^*| z,{\bm{p}}_{-}
+{\mbox{\boldmath{$\kappa$}}}_1 +{\mbox{\boldmath{$\kappa$}}}_3
\rangle - \langle \gamma^*| z,{\bm{p}}_{-}
+{\mbox{\boldmath{$\kappa$}}}_1
+{\mbox{\boldmath{$\kappa$}}}_3+{\mbox{\boldmath{$\kappa$}}}
\rangle
\right\} \nonumber\\
\approx \left|\langle \gamma^*|
z,{\mbox{\boldmath{$\kappa$}}}_3\rangle - \langle \gamma^*|
z,{\mbox{\boldmath{$\kappa$}}}_3 +{\mbox{\boldmath{$\kappa$}}}
\rangle \right|^2\, .
\label{eq:4.17}
\end{eqnarray}
The principal point is that the minijet-minijet inclusive cross
section would depend neither on the minijet momentum nor on the
decorrelation momentum. This proves a complete disappearance of
the azimuthal correlation of minijets with a transverse momentum
below the saturation scale.


\section{ Nonlinear $k_{\perp}$-factorization for the breakup of pions
into forward dijets on nuclei}

\subsection{From pointlike photons to non-pointlike pions}

By requiring the production of forward dijets which satisfy the
$x_{\gamma}=1$ criterion we select the breakup of the $q\bar{q}$
Fock state of the projectile. Above we concentrated on the breakup
of a pointlike projectile, in which case the back-to-back dijets
stem from ``lifting onto  mass-shell'' of $q\bar{q}$ states with a
large intrinsic transverse momentum ${\bm{p}}_{\perp}$. In this
section we go to another extreme case - the non-pointlike
projectile, a pion, with a soft $q\bar{q}$ WF such that the
intrinsic transverse momentum of the quark and antiquark is
limited.

An important point is the r\^ole that unitarity plays in the
isolation of truly inelastic collisions which is effected by a
subtraction of the coherent diffractive components in
(\ref{eq:4.3}). As is well known, in collisions of strongly
interaction hadrons (pions) with opaque nuclei unitarity entails
that coherent elastic scattering off a nucleus, $\pi A \to \pi A$,
makes 50\% of the total hadron-nucleus cross section. On the other
hand, for weakly interacting pointlike projectiles (photons)  it
is coherent diffractive excitation of the continuum $q\bar{q}$
states, $\gamma^* A \to (q\bar{q})~A$, which makes 50\% of the
total DIS cross section \cite{NZZdiffr}. In our formalism as
exposed in section 3, we explicitly associate the subtracted
coherent diffractive component in (\ref{eq:3.1}) with the
continuum $q\bar{q}$ dijets. Strictly speaking, in the case of
incident pions the subtraction of coherent diffractive processes
must include a sum over elastic pions, diffractively excited meson
resonances and, finally, the continuum diffractive states.
Although the quark-antiquark interaction in the $q\bar{q}$ Fock
states of elastic pions and diffractive meson resonances is
important for the formation of pions and its excitations, the
contribution to their wave function from large invariant mass,
$M_{q\bar{q}}$, Fock states vanishes rapidly at very large
$M_{q\bar{q}}$. Then, as argued in \cite{NSSdijet}, the FS
interaction between the FS quark and antiquark can be neglected
and the $q\bar{q}$ plane-wave approximation becomes applicable as
soon as the invariant mass of the forward dijet system exceeds a
typical mass scale of prominent meson resonances. Consequently,
the technique of section 3 and the derivation of a nonlinear
nuclear $k_{\perp}$-factorization in section 4 are fully
applicable to breakup of pions into high-mass hard dijets.


\subsection{In-volume breakup is hard pQCD tractable, probes
the pion distribution amplitude but is subleading at large
${\bm{p}}_{\pm}$}

Whether the breakup of non-pointlike pions into hard dijets,
${\bm{p}}_{\pm}^2 \agt Q_A^2$, is under full control of
perturbative QCD or not needs further scrutiny. In the case of DIS
it was important that the wave function of the photon was a slow
function compared to nuclear WW glue, see the discussion preceding
the derivation of eq.~(\ref{eq:4.16}) in section 4. Here we notice
that for extended heavy nuclei the saturation scale $Q_A$ is much
larger than the ($z$-dependent) intrinsic transverse momentum of
quarks, $Q_{\pi}$, in the pion, so that the pion wave function
would be the steepest function of transverse momentum in the
problem. A closer inspection of the nonlinear
$k_{\perp}$-factorization formula (\ref{eq:4.13}) shows that one
must compare the momentum dependence of the pion WF $\langle
\pi|{\mbox{\boldmath{$\kappa$}}} \rangle$ with that of
$\Phi((1-\beta)\nu_A({\bm{b}}),{\mbox{\boldmath{$\kappa$}}})$ and
or $\Phi(\beta\nu_A({\bm{b}}),{\mbox{\boldmath{$\kappa$}}})$, i.e,
$Q_{\pi}^2$ must be compared to the $\beta$-dependent saturation
scale $Q_{\beta}^2 = \beta Q_{A}^2$ or $ (1-\beta) Q_A^2$.

We shall consider first the contribution to the dijet cross
section from $Q_{\beta}^2 \agt Q_{\pi}^2$, i.e., $\beta_{min}=
Q_{\pi}/Q_A^2 \alt \beta \alt \beta_{\max}= 1-\beta_{min}$. It
describes the in-volume breakup of pions, see the interpretation
of eq.~(\ref{eq:4.2}). Here the pion WF is the steepest one
compared to other factors in (\ref{eq:4.13}) and can be
approximated by a $\delta$-function in transverse momentum space ,
\begin{eqnarray}
\langle z,{\bm{p}} |\pi \rangle   =
(2\pi)^2\delta({\bm{p}}) \int {d^2{\bm{k}} \over (2\pi)^2} \langle
z,{\bm{k}} |\pi \rangle   = (2\pi)^2\delta({\bm{p}})\sqrt{{\pi
\over 2N_c}}F_{\pi}\varphi_{\pi} (Q_\beta^2,z)\, ,
\label{eq:5.1.1}
\end{eqnarray}
which gives
\begin{eqnarray}
{1\over (2\pi)^2} \int
d^2{\mbox{\boldmath{$\kappa$}}}_1
\Phi((1-\beta)\nu_A({\bm{b}}),{\mbox{\boldmath{$\kappa$}}}_1)
\left\{\langle z,{\bm{p}}_{-} +{\mbox{\boldmath{$\kappa$}}}_1
+{\mbox{\boldmath{$\kappa$}}}_3 |\pi\rangle - \langle
z,{\bm{p}}_{-} +{\mbox{\boldmath{$\kappa$}}}_1
+{\mbox{\boldmath{$\kappa$}}}_3+{\mbox{\boldmath{$\kappa$}}}
|\pi\rangle
\right\} \nonumber\\
\approx  \sqrt{{\pi \over 2N_c}}F_{\pi}\varphi_{\pi}
(Q_\beta^2,z)[\Phi((1-\beta)\nu_A({\bm{b}}),{\bm{p}}_-
+{\mbox{\boldmath{$\kappa$}}}_3)
-\Phi((1-\beta)\nu_A({\bm{b}}),{\bm{p}}_-
+{\mbox{\boldmath{$\kappa$}}}_3 +{\mbox{\boldmath{$\kappa$}}})]\,
, \label{eq:5.1.2}
\end{eqnarray}
where we indicated explicitly the factorization scale  $Q_\beta^2$
in the $\pi$DA $\varphi_{\pi}(Q_\beta^2,z)$. Of the two possible
helicity structures in (\ref{eq:2.13}) only the one, $\propto
\delta_{\lambda -\bar{\lambda}}$, related to the pion decay
constant, contributes in this hard regime, see also the related
discussion of diffractive dijets in \cite{NSSdijet}. Then this
contribution to the breakup of pions into high-mass dijets can be
presented in two equivalent forms, which only differ by a
reshuffling of the large jet momentum ${\bm{p}}_-$ between
different factors in the integrand:
\begin{eqnarray}
&&{d\sigma_{\pi} \over d^2{\bm{b}} dz d^2{\bm{p}}_{-}
d^2{\bm{\Delta}}} = {\pi^3 \over N_c} \cdot \alpha_S \sigma_0(x)
T({\bm{b}}) F_{\pi}^2
\nonumber\\
&&\times  \int_{\beta_{min}}^{\beta_{max}} d \beta
\varphi_{\pi}^2(Q_{\beta}^2,z) \int
d^2{\mbox{\boldmath{$\kappa$}}}_3 d^2{\mbox{\boldmath{$\kappa$}}}
f({\mbox{\boldmath{$\kappa$}}})
\Phi(\beta\nu_A({\bm{b}}),{\bm{\Delta}}
-{\mbox{\boldmath{$\kappa$}}}_3 -{\mbox{\boldmath{$\kappa$}}})
\Phi(\beta\nu_A({\bm{b}}),{\mbox{\boldmath{$\kappa$}}}_3)
\nonumber\\
&&\times [\Phi((1-\beta)\nu_A({\bm{b}}),{\bm{p}}_-
+{\mbox{\boldmath{$\kappa$}}}_3)
-\Phi((1-\beta)\nu_A({\bm{b}}),{\bm{p}}_-
+{\mbox{\boldmath{$\kappa$}}}_3 +{\mbox{\boldmath{$\kappa$}}})]^2
\nonumber\\
&&={\pi^3 \over N_c} \cdot \alpha_S \sigma_0(x) T({\bm{b}})
F_{\pi}^2
\nonumber\\
&&\times  \int_{\beta_{min}}^{\beta_{max}} d \beta
\varphi_{\pi}^2(Q_{\beta}^2,z) \int d^2{\bm{q}}
d^2{\mbox{\boldmath{$\kappa$}}} f({\mbox{\boldmath{$\kappa$}}})
\Phi(\beta\nu_A({\bm{b}}),{\bm{p}}_+ -{\bm{q}}
-{\mbox{\boldmath{$\kappa$}}})
\Phi(\beta\nu_A({\bm{b}}),{\bm{p}}_- + {\bm{q}})
\nonumber\\
&&\times [\Phi((1-\beta)\nu_A({\bm{b}}),{\bm{q}})
-\Phi((1-\beta)\nu_A({\bm{b}}),{\bm{q}}+{\mbox{\boldmath{$\kappa$}}})]^2
\, .
\label{eq:5.1.3}
\end{eqnarray}
Alternatively, one could have neglected the ${\bm{r}}$-dependence
of the pion wave function, i.e., put the radial WF $
\Psi_{\pi}(z,{\bm{r}}) \approx \Psi_{\pi}(z,0)$ and proceed with
the calculations which lead to (\ref{eq:4.13}).

At first sight, in close similarity to the breakup into coherent
diffractive dijets \cite{NSSdijet}, the inclusive dijet cross
section is proportional to the $\pi$DA  squared. One must be
careful with the isolation of the leading large-${\bm{p}}_{\pm}$
behavior of the pion breakup, though.

Consider first the former representation of (\ref{eq:5.1.3})
According to \cite{NSSdijet,INDiffGlue}, in the hard region $
f({\mbox{\boldmath{$\kappa$}}}) \sim
1/({\mbox{\boldmath{$\kappa$}}}^2)^{\delta}$ with the exponent
$\delta \sim 2$ and $
\Phi(\nu_A({\bm{b}}),{\mbox{\boldmath{$\kappa$}}}) \approx
\nu_A({\bm{b}})f({\mbox{\boldmath{$\kappa$}}})$ , see
eq.~(\ref{eq:4.9}), so that it is tempting to focus on
${\mbox{\boldmath{$\kappa$}}}_3^2 \alt \beta Q_A^2$ and neglect
${\mbox{\boldmath{$\kappa$}}}_{3}$ compared to ${\bm{p}}_-$ in
(\ref{eq:5.1.2}):
\begin{eqnarray}
[\Phi((1-\beta)\nu_A({\bm{b}}),{\bm{p}}_- -{\mbox{\boldmath{$\kappa$}}}_3)
-\Phi((1-\beta)\nu_A({\bm{b}}),{\bm{p}}_- -{\mbox{\boldmath{$\kappa$}}}_3-{\mbox{\boldmath{$\kappa$}}})]^2 \nonumber\\
\approx [\Phi((1-\beta)\nu_A({\bm{b}}),{\bm{p}}_-)
-\Phi((1-\beta)\nu_A({\bm{b}}),{\bm{p}}_-
-{\mbox{\boldmath{$\kappa$}}})]^{2} \, .
\label{eq:5.1.4}
\end{eqnarray}
Upon taking the convolution (\ref{eq:4.15}), this contribution to
the dijet cross section can be cast into a form reminiscent of
(\ref{eq:4.16}):
\begin{eqnarray}
{d\sigma_{\pi} \over d^2{\bm{b}} dz d^2{\bm{p}}_{-}
d^2{\bm{\Delta}}} &=& T({\bm{b}})\int_ {0}^{\beta_{max}} d \beta
\int d^2{\mbox{\boldmath{$\kappa$}}}
\Phi(2\beta\nu_A({\bm{b}}),{\bm{\Delta}}
-{\mbox{\boldmath{$\kappa$}}}) {d\sigma_{eff} \over dz
d^2{\bm{p}}_+ d^2{\mbox{\boldmath{$\kappa$}}} }\, ,
\label{eq:5.1.5}
\end{eqnarray}
where
\begin{eqnarray} {d\sigma_{eff} \over dz d^2{\bm{p}}_+ d^2{\mbox{\boldmath{$\kappa$}}} } &=&
{1\over 2(2\pi)^2}  \cdot {\pi \over 2N_c} \cdot \alpha_S
\sigma_0(x) F_{\pi}^2 \varphi_{\pi}^2(Q_{\beta}^2,z)
f({\mbox{\boldmath{$\kappa$}}}) \nonumber\\
&\times &[\Phi((1-\beta)\nu_A({\bm{b}}),{\bm{p}}_-)
-\Phi((1-\beta)\nu_A({\bm{b}}),{\bm{p}}_-
-{\mbox{\boldmath{$\kappa$}}})]^2
\label{eq:5.1.6}
\end{eqnarray}
plays the r\^ole of the free-nucleon cross section for DIS,
eq.~(\ref{eq:2.5}) and $\Phi((1-\beta)\nu_A({\bm{b}}),{\bm{p}})$
emerges as the counterpart of the WF $\langle
{\bm{p}}|\gamma^*\rangle$ in (\ref{eq:2.5}), (\ref{eq:4.16}). Now
we notice that this consideration is fully applicable to $\beta <
\beta_{min}$, for which reason we already have put $\beta=0$ for
the lower limit of the $\beta$ integration in (\ref{eq:5.1.5}).

The nuclear WW gluon distribution
$\Phi(2\beta\nu_A({\bm{b}}),{\bm{\Delta}}
-{\mbox{\boldmath{$\kappa$}}})$ in (\ref{eq:5.1.5}) provides a
cutoff of the ${\mbox{\boldmath{$\kappa$}}}$ integration,
${\mbox{\boldmath{$\kappa$}}}^2 \alt \beta Q_A^2 +
{\bm{\Delta}}^2$, which justifies the small-$\kappa$ approximation
\begin{eqnarray}
[\Phi((1-\beta)\nu_A({\bm{b}}),{\bm{p}}_-)
-\Phi((1-\beta)\nu_A({\bm{b}}),{\bm{p}}_- -{\mbox{\boldmath{$\kappa$}}})]^2
\nonumber\\
\approx 2\left( {f({\bm{p}}_-) \over {\bm{p}}_-^2}\right)^2
\delta^2(1-\beta)^2\nu_A^2({\bm{b}}) {\bm{p}}_-^2
{\mbox{\boldmath{$\kappa$}}}^2 \, ,
\label{eq:5.1.7}
\end{eqnarray}
where the azimuthal averaging has been performed. Then the
${\mbox{\boldmath{$\kappa$}}}$ integration subject to
${\mbox{\boldmath{$\kappa$}}}^2 \alt \beta Q_A^2 +
{\bm{\Delta}}^2$ yields
\begin{eqnarray}
\int d^2{\mbox{\boldmath{$\kappa$}}}
f({\mbox{\boldmath{$\kappa$}}}){\mbox{\boldmath{$\kappa$}}}^2
\approx {4\pi^2 \over \sigma_0(x) N_c} G({\bm{\Delta}}^2 + \beta
Q_{A}^{2}) \label{eq:5.1.8}
\end{eqnarray}
and the final results for this contribution to the dijet cross
section reads
\begin{eqnarray}
{d\sigma_{\pi A}^{(volume)} \over d^2{\bm{b}}dz d^2{\bm{p}}_{-}
d^2{\bm{\Delta}}}\Big|_{hard} &\approx& {32\pi^6 F_{\pi}^2
T^3({\bm{b}}) \over N_c^4}\cdot {\delta^2\alpha_S^3({\bm{p}}_-^2)
{\cal
F}^2(x,{\bm{p}}_-^2) \over ({\bm{p}}_-^2)^5} \cdot \nonumber\\
&&\times \int_0^1 d\beta (1-\beta)^2
\varphi_{\pi}^2(Q_{\beta}^2,z) \Phi(2\beta \nu_{A}({\bm{b}}),
{\bm{\Delta}}) G({\bm{\Delta}}^2 + \beta Q_A^2)\, .
\label{eq:5.1.9}
\end{eqnarray}
Here the $\beta$ integration is dominated by the mid-$\beta$
contribution, hence the $\beta$ integration can safely be extended
from 0 to 1. The dominance of the mid-$\beta$ contribution
 means that the incident
pion breaks in the volume of a nucleus, i.e., it selects weakly
attenuating small color-dipole configurations in the incident
pion. For this reason it is uniquely calculable in terms of hard
quantities and collective nuclear WW glue, the auxiliary soft
parameter $\sigma_0(x)$ does not enter (\ref{eq:5.1.9}), hence the
subscript ``hard'' in the l.h.s. of (\ref{eq:5.1.9}). All the
approximations which have lead to the proportionality of the pion
breakup cross section to the $\pi$DA  squared were indeed well
justified. Unfortunately, as far as the $p_+$ dependence is
concerned the in-volume hard absorption gives the subleading,
higher-twist contribution.

\subsection{The leading asymptotics at large ${\bm{p}}_{\pm}$ is
dominated by soft absorption on the front face of a nucleus}

The leading asymptotics at large ${\bm{p}}_{\pm}$ can be isolated
starting with the latter representation of eq.~(\ref{eq:5.1.3}).
It comes from $|{\bm{q}}|=|{\bm{p}}_-
-{\mbox{\boldmath{$\kappa$}}}_3| \ll Q_A$, furthermore, as we
shall see a posteriori the dominant contribution comes from a
still narrower soft domain $|{\bm{q}}| \alt Q_{\pi}$. We start
with the in-volume contribution from $\beta_{min} < \beta < 1
-\beta_{min}$. According to (\ref{eq:4.9}), for such hard jets we
can approximate
\begin{eqnarray}
\Phi(\beta\nu_A({\bm{b}}),{\bm{p}}_+ -{\bm{q}} -{\mbox{\boldmath{$\kappa$}}})
\Phi(\beta\nu_A({\bm{b}}),{\bm{p}}_- + {\bm{q}}) \approx
\beta^2\nu_A^2({\bm{b}})f({\bm{p}}_+)f({\bm{p}}_-)\, .
\label{eq:5.3.1}
\end{eqnarray}
Now, using the convolution identity eq.(\ref{eq:4.15}), we may
simplify
\begin{eqnarray}
&&\int d^2\bm{q} d^2\bm{\kappa} f(\bm{\kappa})
[\Phi((1-\beta)\nu_A({\bm{b}}),{\bm{q}})
-\Phi((1-\beta)\nu_A({\bm{b}}),{\bm{q}}+{\mbox{\boldmath{$\kappa$}}})]^2=
\nonumber \\
&&= 2 \int d^2\bm{\kappa} f(\bm{\kappa})
[\Phi(2(1-\beta)\nu_A({\bm{b}}),{\bm{0}})
-\Phi(2(1-\beta)\nu_A({\bm{b}}),{\bm{\kappa}})] \nonumber \\
&&=- 2 \int d^2\bm{\kappa} f(\bm{\kappa}) \, {\partial \Phi
(2(1-\beta) \nu_A(\bm{b}),\bm{\kappa}^2) \over
\partial\bm{\kappa}^2 }\Big|_{\bm{\kappa}^2=0} \cdot \bm{\kappa}^2
\nonumber \\
&& \simeq {4 \pi \over (1-\beta)^2} \cdot {1 \over Q_A^4} \cdot {1
\over \sigma_0(x) N_c} \cdot G(Q_\beta^2)\, .
\end{eqnarray}
where $Q_\beta^2 = (1-\beta)Q_A^2$ and we made explicit use of the
parametrization (\ref{eq:4.10}). The singular behaviour at $\beta
\to 1$ shows that the inclusive cross section will be dominated by
the soft end-point contribution from $Q_{\beta}^2 \sim Q_{\pi}^2$,
\begin{equation}
\int_{\beta_{min}}^{\beta_{max}} {d\beta G(Q_{\beta}^2) \beta^2
\varphi_\pi^2(Q_\beta^2,z) \over (1-\beta)^2} \sim { G(Q_{\pi}^2)
\varphi_\pi^2(Q_\pi^2,z) \over 1-\beta_{\max}} = {Q_A^2 \over
Q_{\pi}^2} \cdot G(Q_{\pi}^2) \varphi_\pi^2(Q_\pi^2,z) \, ,
\label{eq:5.3.4}
\end{equation}
i.e., the in-volume contribution is squeezed to the breakup of
pions close to the front face of the nucleus. Finally, making use
of (\ref{eq:4.11}) and neglecting the difference between
$G(Q_{\pi}^2)$ and $G(Q_{A}^2)$, we obtain an estimate
\begin{eqnarray}
{d\sigma_{\pi A}^{(soft)} \over d^2{\bm{b}} dz d^2{\bm{p}}_{-}
d^2{\bm{p}}_+} = {4 \pi^4 T^2({\bm{b}}) \alpha_S^2
({\bm{p}}_{\pm}^2)F_{\pi}^2 \over  N_c^3}\cdot
  \phi_{\pi}^2(Q_{\pi}^2,z) \cdot {1\over Q_{\pi}^2}\cdot {{\cal F}(x,{\bm{p}}_+^2)\over
({\bm{p}}_+^2)^2 }\cdot { {\cal F}(x,{\bm{p}}_-^2) \over
({\bm{p}}_-^2)^2}\, .
\label{eq:5.3.5}
\end{eqnarray}
The striking feature of this result is a complete decorrelation of
the two jets. For the explanation of the superscript ``soft'' see
below.

In the evaluation of the contribution from $\beta_{\max} \alt
\beta \alt 1$, i.e., from breakup of pions on the front face of
the nucleus, we can take
$\Phi((1-\beta)\nu_{A}({\bm{b}}),{\mbox{\boldmath{$\kappa$}}}_{1})
= \delta({\mbox{\boldmath{$\kappa$}}}_{1})$, see
eq.~(\ref{eq:4.5}), so that for hard dijets
\begin{eqnarray}
&&\int d^2{\mbox{\boldmath{$\kappa$}}}_1
\Phi((1-\beta)\nu_A({\bm{b}}),{\mbox{\boldmath{$\kappa$}}}_1)
\left\{\langle \pi| z,{\bm{p}}_{-} +{\mbox{\boldmath{$\kappa$}}}_1
+{\mbox{\boldmath{$\kappa$}}}_3 \rangle - \langle \pi |
z,{\bm{p}}_{-} +{\mbox{\boldmath{$\kappa$}}}_1
+{\mbox{\boldmath{$\kappa$}}}_3+{\mbox{\boldmath{$\kappa$}}}
\rangle
\right\} \nonumber\\
&& \approx \langle \pi| z,{\bm{p}}_{-}
+{\mbox{\boldmath{$\kappa$}}}_3 \rangle - \langle \pi |
z,{\bm{p}}_{-}
+{\mbox{\boldmath{$\kappa$}}}_3+{\mbox{\boldmath{$\kappa$}}}
\rangle
\label{eq:5.3.6}
\end{eqnarray}
Putting $\beta=1$ in the rest of the integrand of (\ref{eq:4.13}),
approximating $\int _{\beta_{max}}^1 d\beta \sim Q_{\pi}^2
/Q_{A}^2$, and reshuffling ${\bm{p}}_-$ as in (\ref{eq:5.1.3}) we
obtain
\begin{eqnarray}
{d\sigma^{surface}_{\pi} \over d^2{\bm{b}} dz d^2{\bm{p}}_{-}
d^2{\bm{\Delta}}} &\approx & {1\over 2(2\pi)^2 Q_{A}^2}  \cdot
\alpha_S \sigma_0(x)
T({\bm{b}})\Phi(\nu_A({\bm{b}}),{\bm{p}}_+)\Phi(\nu_A({\bm{b}}),{\bm{p}}_-)
\nonumber\\
&\times & Q_{\pi}^2 \int d^2{\bm{q}}
d^2{\mbox{\boldmath{$\kappa$}}} f({\mbox{\boldmath{$\kappa$}}})
\left|\langle \pi |z,{\bm{q}}\rangle - \langle \pi |z,{\bm{q}} +
{\mbox{\boldmath{$\kappa$}}} \rangle\right|^2\, ,
\label{eq:5.3.7}
\end{eqnarray}
Although $f({\mbox{\boldmath{$\kappa$}}})$ and the pion WF are
distributions of comparable width, for the estimation purposes we
can expand
\begin{eqnarray}
\left|\langle \pi |z,{\bm{q}}\rangle - \langle \pi
|z,{\bm{q}} + {\mbox{\boldmath{$\kappa$}}} \rangle\right|^2 \sim
{\left|\langle \pi |z,{\bm{q}}\rangle \right|^2 \over
(Q_{\pi}^2+{\bm{q}}^2)^2 } 2 {\bm{q}}^2
{\mbox{\boldmath{$\kappa$}}}^2\,.
\label{eq:5.3.8}
\end{eqnarray}
The ${\mbox{\boldmath{$\kappa$}}}$ integration will yield $
G(Q_{\pi}^2+{\bm{q}}^2)\sim G(Q_{\pi}^2)$, see
eq.~(\ref{eq:5.1.8}), whereas
\begin{eqnarray}
2Q_{\pi}^2\int
d^2{\bm{q}} { \left|\langle \pi |z,{\bm{q}}\rangle \right|^2
{\bm{q}}^2 \over (Q_{\pi}^2+{\bm{q}}^2)^2 } \sim {1\over \pi
Q_{\pi}^2} \left| \int d^2{\bm{q}} \langle \pi |z,{\bm{q}}\rangle
\right|^2 \sim {(2\pi)^4\over Q_{\pi}^2}\cdot {1 \over 2N_c}
F_{\pi}^2 \varphi_{\pi}^2( Q_{\pi}^2,z)\, .
\label{eq:5.3.9}
\end{eqnarray}
Notice, that both helicity components of the pion WF would
contribute in this soft regime. Then, our result for
(\ref{eq:5.3.7}) is identical to (\ref{eq:5.3.5}), so that the
final estimate for the soft cross section is (\ref{eq:5.3.5})
times a factor $\sim$ 2.

Several comments on our result (\ref{eq:5.3.5}), (\ref{eq:5.3.7})
are in order. First, the representation (\ref{eq:4.3}) for the
distortion factor entails that the dominance by the contribution
from $\beta \to 1$ corresponds to an absorption of the pion in the
state with normal hadronic size on the front surface of the
nucleus, i.e., it is a soft absorption driven contribution to
excitation of hard dijets, hence the superscript ``soft'' in the
l.h.s. of (\ref{eq:5.3.5}). Second, the same point about absorbed
pions being in the state with large hadronic size is manifest from
the emergence of the soft factor $1/Q_{\pi}^2$ in the dijet cross
section. Third, the back-to-back correlated hard contribution
(\ref{eq:5.1.9}) is suppressed compared to the soft contribution
(\ref{eq:5.3.5}), (\ref{eq:5.3.7}) by a factor
$1/{\bm{p}}_{\pm}^2$, i.e., it has the form of a higher twist
correction. Fourth, as far as the dependence on transverse
momentum is concerned the dijet cross section from the soft
absorption mechanism has the form of a product of two single-jet
cross sections (\ref{eq:2.12}). This property indicates that hard
jets acquire their large transverse momenta from hard scattering
on different nucleons which explains why the transverse momenta of
the quark and antiquark are fully decorrelated both azimuthally
and longitudinally, i.e., in the magnitude of the momenta
$|{\bm{p}}_+|$ and $|{\bm{p}}_-|$ (in the scattering plane). This
must be contrasted to the breakup of photons when the large
momentum of jets comes from the intrinsic momentum in the photon
WF and jets are produced predominantly back-to-back with the scale
for both the azimuthal (out-of-plane) and longitudinal (in-plane)
decorrelations being set by $Q_A$. Consequently, the out-of-plane
decorrelation momentum in the breakup of pions into forward dijets
is predicted to be much larger than in the breakup of photons in
photoproduction or DIS.


\subsection{Transverse energy  associated with dijets
in the photon and pion breakup}

There is still another interesting difference between the breakup
of pointlike photons and non-pointlike pions. It is the surplus
transverse energy of secondary particles associated with the
forward hard jets.

What counterbalances the large transverse momenta ${\bm{p}}_{\pm}$
of the two uncorrelated hard jets? According to
\cite{NSSdijet,Nonlinear}, the hard tail (\ref{eq:4.9}) of the
collective nuclear WW glue which enters (\ref{eq:5.3.1}) is
dominated by a single hard gluon. Then in the pQCD Born
approximation the forward $q,\bar{q}$ hard jets would recoil
against the valence quarks of nucleons of the target nucleus. With
allowance for the QCD evolution effects the recoil is against the
mid-rapidity gluons and quarks, which are separated from the
forward hard dijets by at least several units of rapidity.
Although the partons which counterbalance the forward dijets are
not localized in the rapidity, their overall contribution to the
transverse energy production in an event (with the transverse
energy from forward dijets excluded) will be an amount of $\Delta
E_T \sim |{\bm{p}}_+| +|{\bm{p}}_-|$ in excess of the average
transverse energy in a minimal bias event without hard forward
jets. This surplus transverse energy production $\Delta E_T$ would
not depend on the azimuthal angle between the two jets.

In the breakup of photons the surplus transverse energy $\Delta
E_T$ will be much smaller. Indeed, the large transverse momenta of
jets come from the large intrinsic transverse momentum of the
quark and antiquark in the photon. The dijet recoils against other
secondary particles with a transverse momentum which is precisely
equal to the acoplanarity momentum $|{\bm{\Delta}}|$. The strong
decorrelation, ${\bm{\Delta}}^2 \agt Q_A^2$, is driven by exchange
of one hard gluon, see the discussion of (\ref{eq:4.9}), and we
expect a surplus transverse energy  $\Delta E_T \approx
|{\bm{\Delta}}|$. In the back-to-back configuration,
${\bm{\Delta}}^2 \alt Q_A^2$, the nuclear glue
$\Phi(\nu_A({\bm{b}}),{\bm{\Delta}})$ is a result of the fusion of
$\propto A^{1/3}$ soft gluons, and we expect the surplus
transverse energy  $\Delta E_T \approx Q_A$.


\subsection{Is the pion distribution amplitude measurable in
the pion breakup into dijets?}

The emergence of the $z$-dependent soft factor $1/Q_{\pi}^2$ in
(\ref{eq:5.3.5}), which  depends on the model for soft wave
function of the pion, (see also the analysis
(\ref{eq:5.3.7})-(\ref{eq:5.3.9})), is  an unfortunate
circumstance. It makes the relationship between the $z$-dependence
of the dijet cross section and the $\pi$DA squared a model
dependent one. Still, the experimental isolation of the hard
component (\ref{eq:5.1.9}) and thereby the measurement of the
$\pi$DA is not an entirely impossible task.

The point is that the hard component (\ref{eq:5.1.9}) from the
in-volume breakup gives rise to back-to-back jets within a small
angular cone limited by the decorrelation momentum
${\bm{\Delta}}^2 \alt Q_A^2$. In the ${\bm{\Delta}}$-plane it is a
well defined peak. Consequently, although the $z$-dependence of
$Q_{\pi}^2$ is not under good theoretical control, it can be
determined experimentally measuring the dijet cross section beyond
the back-to-back cone. Then the soft contribution can reliably be
extrapolated into the back-to-back cone and the observed excess
signal can be identified with the hard contribution. Approximating
the hard cross section at ${\bm{\Delta}}\sim 0$ by the integrand
of (\ref{eq:5.1.9}) at $\beta \sim 1/2$ and comparing it to the
soft cross section (\ref{eq:5.3.5}) times the above estimated
factor of three to four, we find
\begin{equation}
{d\sigma^{(hard)} \over d\sigma^{(soft})} \sim
\delta^2 {Q_{\pi}^2 \over \pi {\bm{p}}_+^2}
\label{eq:5.5.1}
\end{equation}
As an example of the model estimate \cite{NSSdijet} for the pion
WF which reproduces the pion electromagnetic form factor, the
$\pi^0 \to 2\gamma$ decay width and the form factor of $\gamma^*
\gamma \pi$ transition, we cite $Q_{\pi}^2 \sim 0.17$ GeV$^2$. The
extraction of the small hard signal is facilitated by its specific
dependence on ${\bm{\Delta}}$.


\subsection{Dijets for the power-law wave function of the pion }

Above we saw how substantially the dijet cross section changes
from the pointlike photon to the non-pointlike pion with the
limited intrinsic transverse momentum of the quark and antiquark
in the pion. It is interesting to see how our main conclusions for
the pion breakup will change -- if at all -- for a power-law wave
function of the form
\begin{eqnarray}
\langle z,
{\bm{p}}|\pi\rangle \propto (2\pi)^2 F_{\pi}\varphi(z)\sqrt{ {\pi
\over 2N_c}}\cdot {1\over \pi} \cdot { Q_{\pi}^2 \over ({\bm{p}}^2
+ Q_{\pi}^2)^2}\, .~~~
\label{eq:5.6.1}
\end{eqnarray}
A detailed discussion of properties of the pion for such a dipole,
Coulomb-like, WF and its applications to the pion form factor and
forward and non-forward parton distributions in the pion is found
in \cite{Radyushkin2}, an early discussion of some of these issues
for the power-law WF of the proton is found in \cite{BGNPZshad}. A
simple choice \cite{Terentiev,BGNPZshad,VM} suggested by the
relativization of Coulomb-like wave functions is $Q_{\pi}^2 =
z(1-z)\Lambda_{\pi}^2 +m_f^2$, cf. eq. (\ref{eq:2.8}) for the
photon, for a slightly different parameterization see
\cite{Radyushkin2}. The dipole WF can be regarded as a minimal
non-pointlike departure from the pointlike pion which would
correspond to a monopole wave function, cf. eq.~(\ref{eq:2.7}).

For the purposes of our discussion, the dipole wave function has
the same asymptotics at large transverse momenta as the
unintegrated nuclear glue (\ref{eq:4.10}). Consequently, following
the discussion in \cite{Nonlinear,LIYaF} the convolution
(\ref{eq:5.1.2}) can be evaluated as
\begin{eqnarray}
{1\over
(2\pi)^2} \int d^2{\mbox{\boldmath{$\kappa$}}}_1
\Phi((1-\beta)\nu_A({\bm{b}}),{\mbox{\boldmath{$\kappa$}}}_1)
\left\{\langle \pi| z,{\bm{p}}_{-} +{\mbox{\boldmath{$\kappa$}}}_1
+{\mbox{\boldmath{$\kappa$}}}_3 \rangle - \langle \pi |
z,{\bm{p}}_{-} +{\mbox{\boldmath{$\kappa$}}}_1
+{\mbox{\boldmath{$\kappa$}}}_3+{\mbox{\boldmath{$\kappa$}}}
\rangle
\right\} \nonumber\\
\approx  \sqrt{{\pi \over 2N_c}}F_{\pi}\varphi_{\pi}
(Q_\beta^2,z)[\tilde{\Phi}(Q_{\beta}^2({\bm{b}})+Q_{\pi}^2,{\bm{p}}_-
+{\mbox{\boldmath{$\kappa$}}}_3)
-\tilde{\Phi}(Q_{\beta}^2({\bm{b}})+Q_{\pi}^2,{\bm{p}}_-
+{\mbox{\boldmath{$\kappa$}}}_3 +{\mbox{\boldmath{$\kappa$}}})]
\label{eq:5.6.2}
\end{eqnarray}
where
$\tilde{\Phi}(Q_{\beta}^2({\bm{b}})+Q_{\pi}^2,{\mbox{\boldmath{$\kappa$}}})$
is described by the parameterization (\ref{eq:4.10}) subject to
the substitution
\begin{eqnarray}
Q_A^2 \to Q_{\beta}^2({\bm{b}})+Q_{\pi}^2\,.
\label{eq:5.6.3}
\end{eqnarray}
The separation of the leading large-${\bm{p}}_{\pm}$ asymptotics
and isolation of the hard contribution for the dipole Ansatz fully
corroborates the main conclusions of sections 5.2 and 5.3 on the
dominance of the in-volume contribution to the back-to-back
correlated hard dijets and the surface breakup contribution into
uncorrelated dijets. The model dependence enters through
$Q_{\pi}^2$ in (\ref{eq:5.6.3}) and can be neglected at large
$\beta$, i.e., for the in-volume breakup of pions. However, the
model dependent $Q_{\pi}^2$ would dominate (\ref{eq:5.6.3}) for
breakup on the front surface of the nucleus.

\section*{Summary and conclusions}

We presented a comparison of consequences of the nuclear
$k_{\perp}$-factorization for the breakup of non-pointlike pions
and pointlike photons into forward dijets. In striking contrast to
the pQCD tractable hard breakup of photons, the dominant
contribution to the breakup of pions starts from the soft breakup
of pions into quark and antiquark at the front face of the target
nucleus followed by hard intranuclear rescattering of the quark
and antiquark. The most striking prediction is a complete
azimuthal decorrelation of hard jets in the breakup of pions. An
obvious implication is that the out-of-plane dijet decorrelation
momentum squared $\langle \Delta_T^2 \rangle_{\pi A}$ in the
breakup of pions must be much larger than $\langle \Delta_T^2
\rangle_{\gamma A}$ in the breakup of photons.

A direct comparison of the  A-dependence of photo- and
pion-produced dijets at $\sqrt{s}=21$ GeV has been performed in
the E683 Fermi-lab experiment \cite{E683} and gave a solid
evidence for $\langle \Delta_T^2 \rangle_{\pi A} > \langle
\Delta_T^2 \rangle_{\gamma A}$. Unfortunately, these data are on
mid-rapidity jets at relatively large values of $x \agt x_A$,
beyond the applicability of the concept of fusion of partons.
Coherent diffractive forward dijets have been observed at
Fermi-lab by the E791 collaboration \cite{E791}. The experimental
identification of diffractive dijets in the E791 experiment has
been facilitated by their exact back-to-back property,
${\bm{\Delta}}^2 \alt 1/R_A^2$. A similar identification of
forward dijets from the breakup of pions in inelastic $\pi A$
collisions might be problematic, but our principal prediction of
$\langle \Delta_T^2 \rangle_{\pi A} > \langle \Delta_T^2
\rangle_{\gamma A}$ can be tested even under the most liberal
event selection criteria.

Specifically, one only needs to study the azimuthal distribution
properties of the hadronic subsystem in the first several units of
the forward rapidity. For instance, one can define on an
event-by-event basis the two-dimensional analogs of the familiar
thrust and sphericity variables (for the review see \cite{SLWu}).
Then we predict that the forward system in $\pi A$ interactions
will be more spherical than in $\gamma A$ interactions, and the 2D
thrust for  $\gamma A$ with be larger than for $\pi A$. Such an
analysis can be performed also in the COMPASS experiment at CERN
in which the forward systems produced by photons and pions can be
studied in the same apparatus \cite{COMPASS}. Still another
observable which differentiates between the breakup of pointlike
photons and non-pointlike pions is a surplus transverse energy of
secondary particles associated with forward hard jets - for the
same transverse momenta of jets it is larger in $\pi A$ than in
$\gamma A$ collisions.

The experimental isolation of the higher-twist back-to-back
correlated hard contribution from the in-volume breakup is
feasible and would allow the determination of the pion
distribution amplitude. This challenging task can be accomplished
because the background from decorrelated dijets can be determined
experimentally and the back-to-back contribution has a specific
dependence on the decorrelation momentum which broadens with the
target mass number.

The breakup of pions into hard dijets involves the stage of soft
absorption at the front face of the target nucleus and the cross
section is sensitive to models of the pion wave function.
Nonetheless, the predicted transverse momentum dependence of
pionic dijets does not change from soft models with a strong bound
on the intrinsic momentum of quarks in the pion to modern
semihard, power-law (dipole) wave functions.

\bigskip

We are glad to contribute to the Yu.A. Simonov Festschrift. One of
the authors (NNN) recalls vividly from his student years, late
60's, how much I.S. Shapiro, then the leader of the Nuclear Theory
Group, praised Simonov's work on the hyperspherical approach to
many--body systems, the subject of Yurii Antonovich's habilitation
thesis. Ever since then Yurii Antonovich remained at the forefront
of hadron physics and non-perturbative QCD and we take this
occasion to wish Yurii Antonovich to keep his vigour and
creativity.

\bigskip

This work has been partly supported by the INTAS grant 00-00366
and the DFG grant 436RUS17/72/03
\pagebreak\\

\newpage

\begin{figure*}[t!]
\setcaptionmargin{5mm}
\onelinecaptionsfalse 
\includegraphics[width=12cm,height=14cm]{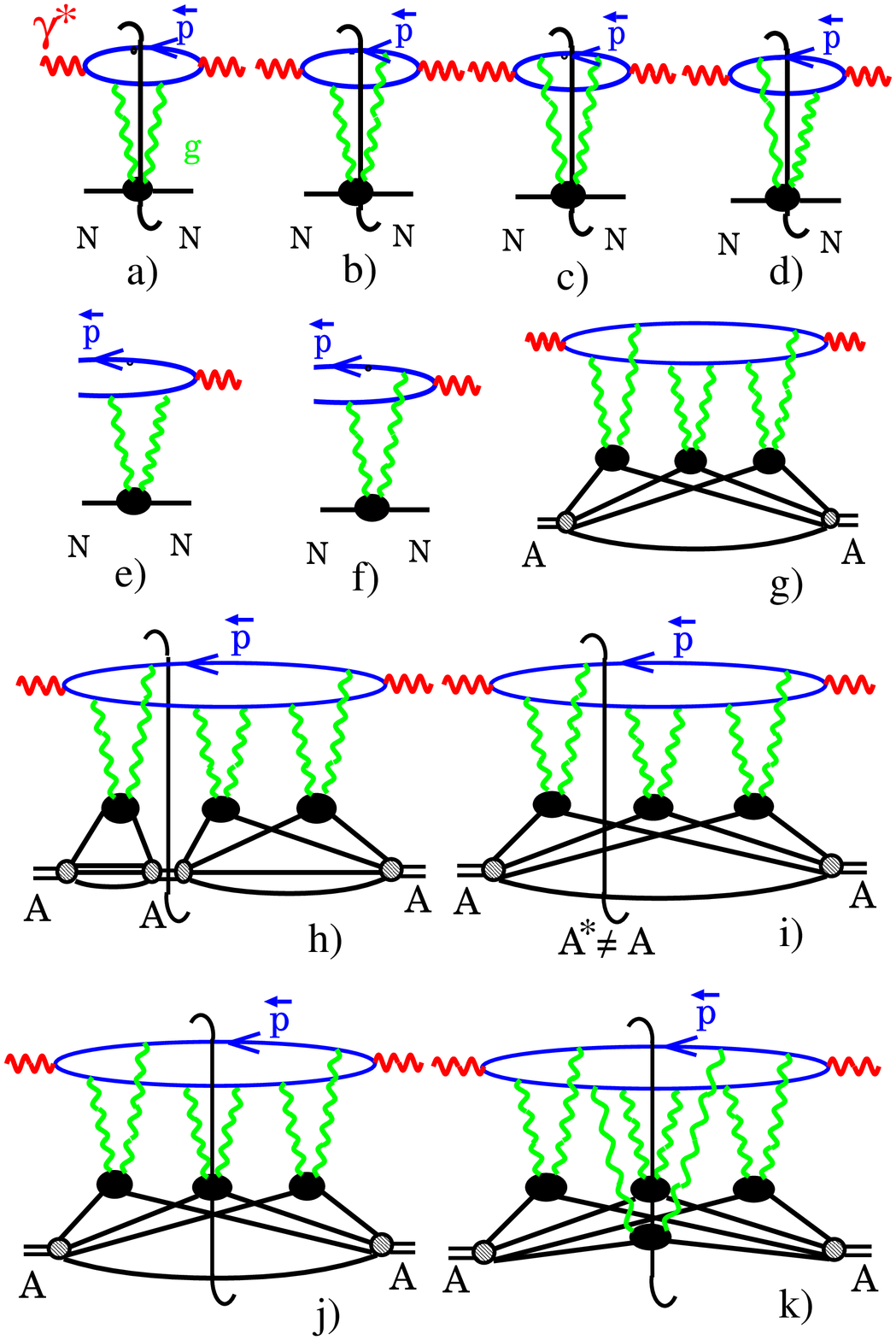}
\captionstyle{normal}
\caption{The pQCD diagrams for inclusive (a-d) and diffractive
(e,f) DIS off protons and nuclei (g-k). Diagrams (a-d) show the
unitarity cuts with color excitation of the target nucleon, (g) -
a generic multiple scattering diagram for Compton scattering off
nucleus, (h) - the unitarity cut for a coherent diffractive DIS,
(i) - the unitarity cut for quasielastic diffractive DIS with
excitation of the nucleus $A^*$, (j,k) - the unitarity cuts for
truly inelastic DIS with single and multiple color excitation of
nucleons of the nucleus.}
\end{figure*}
\begin{figure*}[t!]
\setcaptionmargin{5mm}
\onelinecaptionsfalse 
\includegraphics[width=12cm,height=10cm]{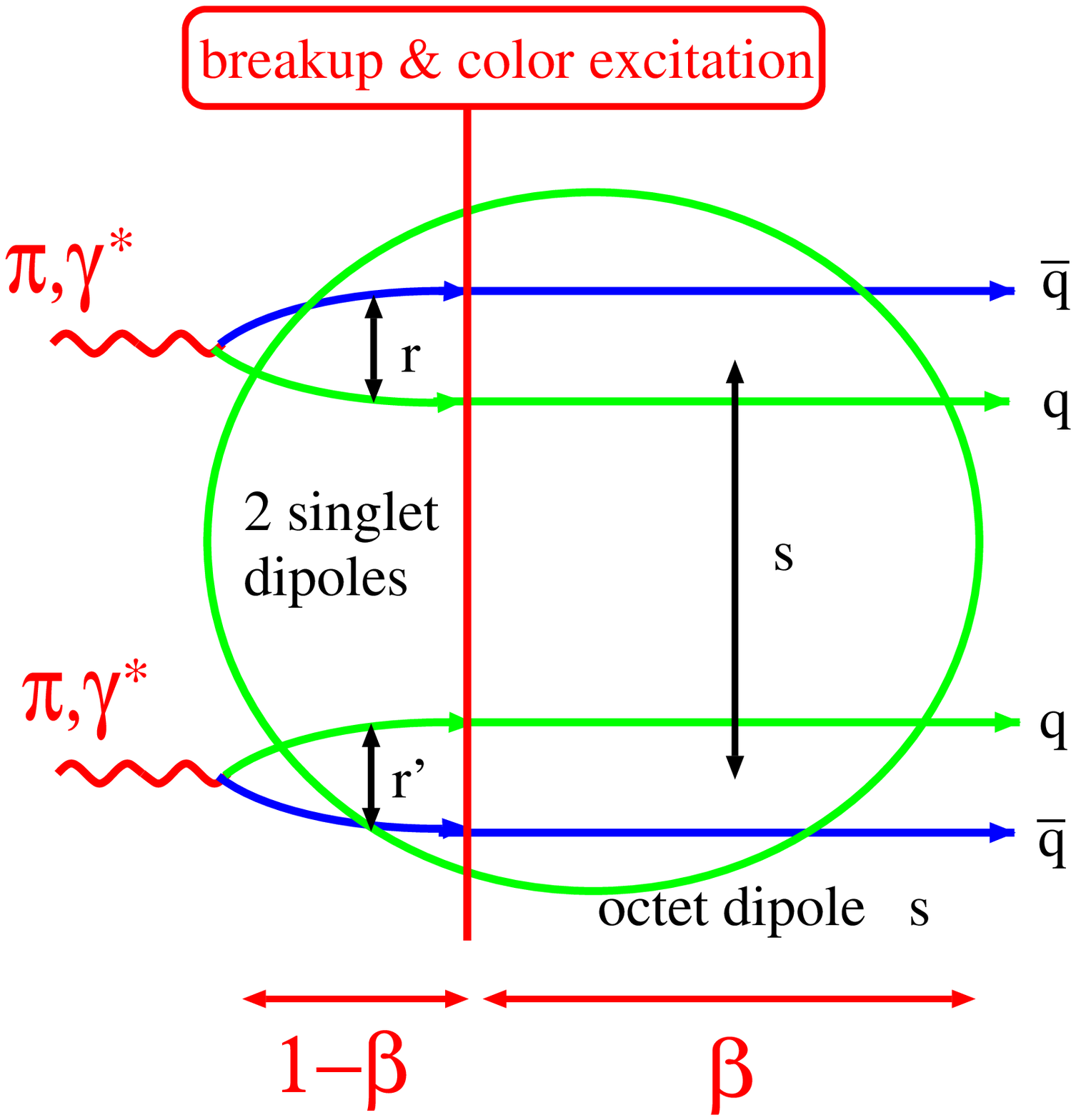}
\captionstyle{normal}
\caption{To the leading order the dijet system is excited into the
color octet state on one nucleon after having travelled a fraction
$1-\beta$ of the nucleus. After that an almost pointlike octet
$q\bar{q}$--system travels the remaining fraction $\beta$.}
\end{figure*}

\begin{thebibliography}{299}

\bibitem{Textbook}
\refitem{book} E. Leader and E. Predazzi, \emph{Introduction to
Gauge Theories and Modern Particle Physics, v.1}, Cambridge
University Press, Cambridge, 1996; \refitem{book} G. Sterman,
\emph{ An Introduction to Quantum Field Theory}, Cambridge
University Press, Cambridge, 1993.

\bibitem{Mueller1}
\refitem{article} A.H. Mueller,  Nucl. Phys. B {\bf 335} (1990)
115.

\bibitem{Mueller} 
\refitem{article}
 A.H. Mueller,  Nucl. Phys. B {\bf 558}
(1999) 285; \refitem{report;prevau}
in {\emph{QCD PERSPECTIVES ON HOT AND DENSE MATTER: Proceedings.}}
Edited by J.-P. Blaizot and E. Iancu. Dordrecht, The Netherlands,
Kluwer, (2002) (NATO Science Series, II, Mathematics, Physics, and
Chemistry, Vol. 87), pp. 45-72, [arXiv: hep-ph/0111244].

\bibitem{McLerran}
\refitem{article} L. McLerran and R. Venugopalan, Phys. Rev. D
{\bf 49} (1994) 2233; \refitem{article} J.Jalilian-Marian, A.
Kovner, L.D. McLerran and H. Weigert, Phys. Rev. D {\bf 55} (1997)
5414; \refitem{report} E. Iancu, A. Leonidov and L. McLerran,
in {\emph{QCD PERSPECTIVES ON HOT AND DENSE MATTER: Proceedings.}}
Edited by J.-P. Blaizot and E. Iancu. Dordrecht, The Netherlands,
Kluwer, (2002) (NATO Science Series, II, Mathematics, Physics, and
Chemistry, Vol. 87), pp. 73-145, [arXiv: hep-ph/0202270].

\bibitem{Saturation}
\refitem{article} N.N. Nikolaev, W. Sch\"afer, B.G. Zakharov and
V.R. Zoller,
JETP Lett.\  {\bf 76} (2002) 195 [Pisma Zh.\ Eksp.\ Teor.\ Fiz.\
{\bf 76} (2002) 231]

\bibitem{Nonlinear}
\refitem{article}
 N.N. Nikolaev, W. Sch\"afer, B.G. Zakharov and V.R. Zoller,
 J.\ Exp.\ Theor.\ Phys.\  {\bf 97} (2003) 441
[Zh.\ Eksp.\ Teor.\ Fiz.\  {\bf 124} (2003) 491]

\bibitem{LIYaF}
\refitem{report}
  I.P. Ivanov, N.N. Nikolaev, W.
Sch\"afer, B.G. Zakharov and V.R. Zoller, Lectures on Diffraction
and Saturation of Nuclear Partons in DIS off Heavy Nuclei,
Proceedings of 36-th Annual Winter School on Nuclear and Particle
Physics and 8-th St. Petersburg School on Theoretical Physics, St.
Petersburg, Russia, 25 Feb - 3 Mar 2002. [arXiv: hep-ph/021216];
\refitem{report,prevau} High Density QCD, Saturation and
Diffractive DIS. Invited talk at the NATO Advanced Research
Workshop on Diffraction 2002, Alushta, Ukraine, 31 Aug - 6
September 2002, [arXiv: hep-ph/0212176]; \refitem{report,prevau}
Diffractive Hard Dijets and Nuclear Parton Distributions,
Proceedings of the Workshop on Exclusive Processes at High
Momentum Transfer, Jefferson Lab, May 15-18, 2002. Editors A.
Radyushkin and P. Stoler, World Scientific, 2002, pp. 205-213,
[arXiv:hep-ph/0207045]; \refitem{article,prevau} High Density QCD
and Saturation of Nuclear Partons, Proceedings of the Conference
on Quark Nuclear Physics (QNP'2002), June 9-14, J\"ulich, Germany,
editors C. Elster, J. Speth and Th. Walcher, Eur. Phys. J. A {\bf
18} (2003), 437 , [arXiv: hep-ph/0209298]; \refitem{report,prevau}
High Density QCD, Saturation and Diffractive DIS. Plenary talk at
the International Symposium on Multiparticle Dynamics (ISMD'2002),
Alushta, Ukraine, 8-14 September 2002, in {\emph {MULTIPARTICLE
DYNAMICS: ISMD 2002: Proceedings.}} Edited by A. Sissakian, G.
Kozlov, E. Kolganova. River Edge, N.J., World Scientific, 2003,
pp.209-220,[arXiv: hep-ph/0212176].


\bibitem{Azimuth}
\refitem{article}
 A. Szczurek, N.N. Nikolaev, W. Sch\"afer, and J. Speth,
Phys. Lett. B {\bf 500} , 254 (2001)

\bibitem{Forshaw}
\refitem{article} J.~R.~Forshaw and R.~G.~Roberts, Phys.\ Lett.\ B
{\bf 335}, 494 (1994); \refitem{article} A.~J.~Askew, D.~Graudenz,
J.~Kwiecinski and A.~D.~Martin, Phys.\ Lett.\ B {\bf 338}, 92
(1994); \refitem{article} J.~Kwiecinski, A.~D.~Martin and
A.~M.~Stasto, Phys.\ Lett.\ B {\bf 459}, 644 (1999)

\bibitem{NZfusion}
\refitem{article} N.N. Nikolaev and V.I. Zakharov, Sov. J. Nucl.
Phys. {\bf 21} (1975) 227; [Yad. Fiz. {\bf 21} (1975) 434];
\refitem{article} Phys. Lett. B {\bf 55} (1975) 397.

\bibitem{NSSdijet}
\refitem{article} N.N. Nikolaev, W. Sch\"afer and G. Schwiete,
JETP Lett. {\bf 72} (2000) 583; [Pisma Zh. Eksp. Teor. Fiz. {\bf
72} (2000) 583]; \refitem{article;prevau} Phys. Rev. D {\bf 63}
(2001) 014020.

\bibitem{Frullani}
\refitem{article} S. Frullani and J. Mougey, Adv. Nucl. Phys. {\bf
14}, 3 (1984).

\bibitem{PionDA}
\refitem{article} V.L. Chernyak and A.R. Zhitnitsky, Phys. Rept.
{\bf 112} (1984) 173;
\refitem{article}
 G.P. Lepage and S.J. Brodsky, Phys. Rev. D
{\bf 22} (1980) 2157; \refitem{article} S.J. Brodsky, H.-C. Pauli
and S.S. Pinsky, Phys. Rept. {\bf 301} (1998) 299.
\refitem{article} R. Jakob and P. Kroll, Phys. Lett. B {\bf 315}
(1993) 463; Erratum-ibid. B {\bf 319} (1993);
\refitem{article}
A.~P.~Bakulev, S.~V.~Mikhailov and N.~G.~Stefanis,
Phys.\ Lett.\ B {\bf 578} (2004) 91

\bibitem{Regensburg}
\refitem{article} V.M. Braun, D.Yu. Ivanov, A. Sch\"afer, L.
Szymanowski, Phys. Lett. B {\bf 509} (2001) 43;
\refitem{article,prevau} Nucl. Phys. B {\bf 638} (2002) 111.

\bibitem{Chernyak}
\refitem{article} V.L. Chernyak, Phys. Lett. B {\bf 516} (2001)
116; \refitem{article} V.L. Chernyak, A.G. Grozin,  Phys. Lett. B
{\bf 517} (2001) 119.

\bibitem{COMPASS}
\refitem{report} COMPASS. A Proposal for a COmmon Muon and Proton
Apparatus for Structure and Spectroscopy. CERN/SPSLC 96-14,
SPSLC/P297, 1 March 1996.

\bibitem{NZ91} 
\refitem{article} N.N.~Nikolaev and B.G.~Zakharov, Z. Phys. C {\bf
49} (1991) 607

\bibitem{NZ92}  
\refitem{article} N.N. Nikolaev and B.G. Zakharov, Z. Phys. C {\bf
53} (1992) 331.

\bibitem{NZ94} 
\refitem{article} N.N. Nikolaev and B.G. Zakharov,  J. Exp. Theor.
Phys. {\bf 78} (1994) 806; [Zh. Eksp. Teor. Fiz. {\bf 105} (1994)
1498]; \refitem{article;prevau} Z. Phys. C {\bf 64} (1994) 631.

\bibitem{NZZlett}
\refitem{article} N.N. Nikolaev, B.G. Zakharov and V.R. Zoller,
JETP Lett. {\bf 59} (1994) 6

\bibitem{NZglue}
\refitem{article} N.N. Nikolaev and B.G. Zakharov. Phys. Lett. B
{\bf 332} (1994) 184

\bibitem{NZZdiffr}
\refitem{article} N.N. Nikolaev, B.G. Zakharov and V.R. Zoller, Z.
Phys. A {\bf 351} (1995) 435.

\bibitem{BGNPZunit}
\refitem{article} V. Barone, M. Genovese, N.N. Nikolaev, E.
Predazzi and B.G. Zakharov, Phys. Lett. B {\bf 326} (1994) 161

\bibitem{BGNPZshad}
\refitem{article} V. Barone, M. Genovese, N.N. Nikolaev, E.
Predazzi and B.G. Zakharov,  Z. Phys. C {\bf 58} (1993) 541

\bibitem{BFKL}
\refitem{article} L.N. Lipatov, Sov. J. Nucl. Phys {\bf 23} (1976)
338; \refitem{article} E.A. Kuraev, L.N. Lipatov and V.S. Fadin,
Sov. Phys. JETP {\bf 44} (1976) 443; \refitem{article;ibid} {\bf
45} (1977) 199; \refitem{article} Ya.Ya. Balitsky and L.N.
Lipatov, Sov. J. Nucl. Phys {\bf 28} (1978) 822

\bibitem{INDiffGlue}
\refitem{article} I.P. Ivanov and N.N. Nikolaev, Phys. Atom. Nucl.
{\bf 64}, 753 (2001), [ Yad. Fiz. {\bf 64}, 813 (2001)];
\refitem{article;prevau} Phys. Rev. D {\bf 65} 054004 (2002).

\bibitem{Andersson}
\refitem{article} B.~Andersson {\it et al.}  (Small x Collab.),
Eur.\ Phys.\ J.\ C {\bf 25}, 77 (2002)

\bibitem{ggFusion}
\refitem{article} T. Ahmed  et al. (H1 Collab.), Nucl. Phys. B
{\bf 445}, 195 (1995), and references therein.

\bibitem{NZsplit}
\refitem{article} N.N.Nikolaev and B.G.Zakharov,  Phys. Lett. B
{\bf 332} (1994) 177

\bibitem{Kroll}
\refitem{article} M. Diehl, T. Feldmann, R. Jakob and  P. Kroll,
Eur. Phys. J. C {\bf  8 } (1999) 409

\bibitem{Radyushkin}
\refitem{article} A.V. Radyushkin, Phys. Rev. D{\bf 58} (1998)
114008

\bibitem{Jaus}
\refitem{article} W. Jaus, Phys. Rev. D {\bf 44} (1991) 2851

\bibitem{PDG}
\refitem{article} Review of Particle Physics,  Eur. Phys. J. C
{\bf 3} (1998) 1

\bibitem{Glauber}
\refitem{book} R. J. Glauber, in {\emph{ Lectures in Theoretical
Physics,}}
 edited by W. E. Brittin et al. (Interscience Publishers, Inc., New York, 1959), Vol. 1, p. 315.

\bibitem{BGZpositronium}
\refitem{article} B.G. Zakharov, Sov. J. Nucl. Phys. {\bf 46}
(1987) 92; [Yad. Fiz. {\bf 46} (1987) 148].


\bibitem{NPZcharm}
\refitem{article} N.N. Nikolaev, G. Piller and B.G. Zakharov.  J.
Exp. Theor. Phys. {\bf  81} (1995) 851; \refitem{article;prevau}
Z. Phys. A {\bf 354} (1996) 99

\bibitem{LPM}
\refitem{article} B.G. Zakharov, JETP Lett. {\bf 63} (1996) 952;
\refitem{article;prevau} JETP Lett. {\bf 65} (1997) 615;
\refitem{article;prevau} Phys. Atom. Nucl. {\bf 61} (1998) 838;

\bibitem{RHIC_STAR}
\refitem{article}
 C. Adler, et al. (STAR Collab.),
Phys. Rev. Lett. {\bf 90}, 082302 (2003)

\bibitem{Radyushkin2}
\refitem{article} A.~Mukherjee, I.~V.~Musatov, H.~C.~Pauli and
A.~V.~Radyushkin,
Phys.\ Rev.\ D {\bf 67} (2003) 073014.

\bibitem{Terentiev}
\refitem{article} M.V. Terentiev, Sov. J. Nucl. Phys. {\bf 24} 106
(1976); [Yad. Fiz. {\bf 24}, 207 (1976)]

\bibitem{VM}
\refitem{article} J. Nemchik, N.N. Nikolaev and  B.G. Zakharov,
Phys. Lett. B {\bf 341}, 228 (1994); \refitem{article} J. Nemchik,
N.N. Nikolaev, E. Predazzi and B.G. Zakharov, Phys. Lett. B {\bf
374}, 199 (1996); \refitem{article;prevau}  Z. Phys. C {\bf 75},
71 (1997)

\bibitem{E683}
\refitem{article} D. Naples et al. (E683 Collab.), Phys. Rev.
Lett. {\bf 72}, 2341 (1994)

\bibitem{E791}
\refitem{article} E.M. Aitala et al. (E791 Collab.) Phys. Rev.
Lett. {\bf 86}, 4768 (2001)

\bibitem{SLWu}
\refitem{article}
 S.L. Wu, Phys. Rep. {\bf 107}, 59 (1984)


\end{thebibliography}
\end{document}